\documentclass[aps,prb,reprint,unsortedaddress,superscriptaddress]{revtex4-1}

\usepackage{amsmath,amssymb}

\usepackage{graphicx}

\usepackage{bm}

\begin{document}

\title{Finite-size energy gap in weak and strong topological insulators}

\begin{abstract}
The non-trivialness of a topological insulator (TI) is characterized 
either by a bulk topological invariant or by the existence of a protected metallic surface state. 
Yet, in realistic samples of finite size this non-trivialness does not necessarily 
guarantee the {\it gaplessness} of the surface state. 
Depending on the geometry and on the topological indices, 
a finite-size energy gap of different nature can appear, and correspondingly, 
exhibits various scaling behaviors of the gap.
The spin-to-surface locking provides one of such gap-opening mechanisms,
resulting in a power-law scaling of the energy gap.
%\textcolor{blue}{
Weak and strong TI's show different degrees of sensitivity to the geometry of the sample.
As a noteworthy example,
a strong TI nanowire of a rectangular prism shape is shown to be
more gapped than that of a weak TI of precisely the same geometry.
\end{abstract}

\date{\today}

\author{Ken-Ichiro Imura}
\affiliation{Department of Quantum Matter, AdSM, Hiroshima University, Higashi-Hiroshima 739-8530, Japan}
\author{Mayuko Okamoto}
\affiliation{Department of Quantum Matter, AdSM, Hiroshima University, Higashi-Hiroshima 739-8530, Japan}
\author{Yukinori Yoshimura}
\affiliation{Department of Quantum Matter, AdSM, Hiroshima University, Higashi-Hiroshima 739-8530, Japan}
\author{Yositake Takane}
\affiliation{Department of Quantum Matter, AdSM, Hiroshima University, Higashi-Hiroshima 739-8530, Japan}
\author{Tomi Ohtsuki}
\affiliation{Department of Physics, Sophia University, 102-8554 Tokyo, Japan}

\maketitle

\section{Introduction}

The non-trivialness  of a TI is 
often characterized by
the presence of a gapless surface state.
\cite{JMoore_review, HasanKane}
A one-to-one correspondence can be established
between the (non-) trivialness of a bulk topological invariant and
the presence vs. absence of the gapless surface state
(bulk-surface correspondence).
However, precisely speaking,
for such a gapless state to be existent,
both the trivial and non-trivial sides are semi-infinite, 
separated by an infinitely large interface.
The above distinction can be made, therefore, 
only in such an idealized situation.
TI samples, in reality, 
occupy only a finite domain of the space,
and have also a variety of shapes
surrounded generally by a curved or folded surface(s).
%\textcolor{blue}{
In experiments it is also the case that
some TI samples of nanometer scale size
exhibit a clear gapless surface state, while
other samples of the same chemical composition but of a different geometrical shape
do not necessarily exhibit a clear signature of topological non-triviality.
Such an issue will be addressed in this paper.

%\textcolor{blue}{
The main scope of the paper is concomitant with the observation
that there are three different gap-opening mechanisms effective
in the samples of finite size.
The most primitive among them
is the one due to mixing of the surface electronic wave functions 
on the opposing sides, e.g., of an infinitely large slab-shaped sample.
Such an energy gap
associated with the finite thickness of the gapped bulk,
decays exponentially as a function of the thickness of the slab,
and is in practice almost irrelevant 
except in extremely thin film samples.
\cite{thin_film}
The low-energy (surface) electronic spectrum in the slab geometry
suffers indeed only from this type of 
exponentially small finite-size energy gap.
\cite{Shen_PRL, Shen_PRB, Linder, Shen_NJP}
%%%%%%%%%%%%%%%%%%%%%%%%
The second mechanism to open a gap in the surface electronic spectrum,
which is also more relevant in magnitude,
is the so-called ``spin-to-surface locking''.
\cite{Ashvin_PRL, Mirlin_PRL, JMoore_PRL, k2}
The electronic spin in the {\it a priori} gapless
surface state on a curved surface of TI
has a tendency
to be locked in-plane to the local tangent of the surface.
%The spin-to-surface locking can be also regarded as a consequence of (spin) Berry phase of $\pi$.
In the cylindrically symmetric case, 
the spin-to-surface locking results in the
{\it half-integral} quantization of
the orbital angular momentum along the axis of the cylinder.
The half-odd integer quantization gaps out the spectrum, and this gap
decays only algebraically;
qualitatively
more relevant than the gap of the previous type.
The spin-to-surface locking leads, indeed, 
irrespective of the presence of cylindrical symmetry, e.g., in a prism-shaped sample,
to opening of the gap.

Another aspect of the topological insulator 
which we aim at exploring in this paper
is the role of anisotropy,
especially in the  weak topological insulator (WTI) phase.
This is much related to the third mechanism of gap-opening, 
which occurs due to the interplay of the anisotropy of WTI
and the specific geometry we will focus on
(case of the prism-shaped geometry).
%%%%%%%%%%%%%%%%%%%%%%%%%%%%
In three spatial dimensions (3D), $\mathbb Z_2$  topological insulator is known to be 
characterized by four $\mathbb Z_2$  indices, 
\cite{FuKaneMele, MooreBalents, Roy}
the principal (strong) index $\nu_0$ 
and other ``weak'' indices $\nu_1, \nu_2, \nu_3$, instead of a single $\mathbb Z_2$ index 
in the case of 2D. 
The principal index $\nu_0$ is used to distinguish a STI ($\nu_0 =1$) 
from trivial and weak topological insulators ($\nu_0 =0$). 
In a WTI
at least one of its weak indices exhibits a nontrivial value ($=1$).
A WTI shows generally an even number of helical Dirac cones on its surfaces, 
but on the surface normal to its ``weak vector'' $\vec{\nu} = (\nu_1, \nu_2, \nu_3)$ 
it shows no Dirac cone.
The WTI can be viewed as stacked layers of 2D $\mathbb Z_2$  topological insulators.
In this regard the set of weak indices $(\nu_1 \nu_2 \nu_3)$
can be regarded as the Miller index of such stacked layers.
Since gapless surface states are expected to form only at the edge of
the stacked layers, 
one can naturally understand that
no Dirac cone is formed on a surface normal to $\vec{\nu}$
in this picture.
To summarize,
the WTI bears two Dirac cones on surfaces {\it parallel} to $\vec{\nu}$
and no Dirac cone on surfaces {\it normal} to $\vec{\nu}$.
When this characteristic feature is combined with
the specific (rectangular) prism geometry,
the anisotropy of a WTI manifests as 
an alternating size-dependence of the energy gap;
the magnitude of the gap is qualitatively different
whether the number of ``stacked layers'' is
even or odd.
%\textcolor{yellow}{The origin of this even/odd feature is revealed in the last part of the paper.}
%\textcolor{blue}{
It will be demonstrated that
weak and strong TI's show different degrees of sensitivity to the geometry of the sample.

The periodic table of topological insulators and superconductors 
classifies them by the nature of strong indices
characterizing the system.
The weak indices are not shown, at least explicitly on the table.
\cite{TeoKane, Ran_arxiv}
Showing
an even number of Dirac cones on its surfaces, 
WTI is {\it a priori} considered not to be robust.
But recently, a few counter examples to this common belief have been proposed.
One is the existence of protected gapless helical modes along
a dislocation line in the WTI.
\cite{Ran_nphys, k1}
More recently, 
a couple of papers have appeared,
demonstrating that
an even number of Dirac cones on the surface of WTI 
are actually not that fragile against disorder.
\cite{Mong,Stern}
Here, we point out that 
in a specific situation in the prism-shaped geometry
the surface state of a WTI is in a sense
``more strongly protected''
from a finite-size energy gap 
than that of a STI.

The paper is organized as follows.
In Sec. II we introduce our effective model Hamiltonian for
3D anisotropic topological insulators.
The phase diagram of the model is determined by the calculation of
topological numbers in the bulk.
In Sec. III we discuss different origins of
the finite-size energy gap, 
highlighting the role of spin-to-surface locking
in the cylindrical geometry.
In Sec. VI we demonstrate that in the more realistic rectangular-prism geometry,
three types of gap-opening appear and
disappear by a small change of model parameters,
leading to an intricate size dependence of the gap.
Sec. VI is devoted to Conclusions.

%%%%%%%%%%%%%%%%%%%%%%%%%%
\begin{figure}
\begin{center}
\includegraphics[width=8cm]{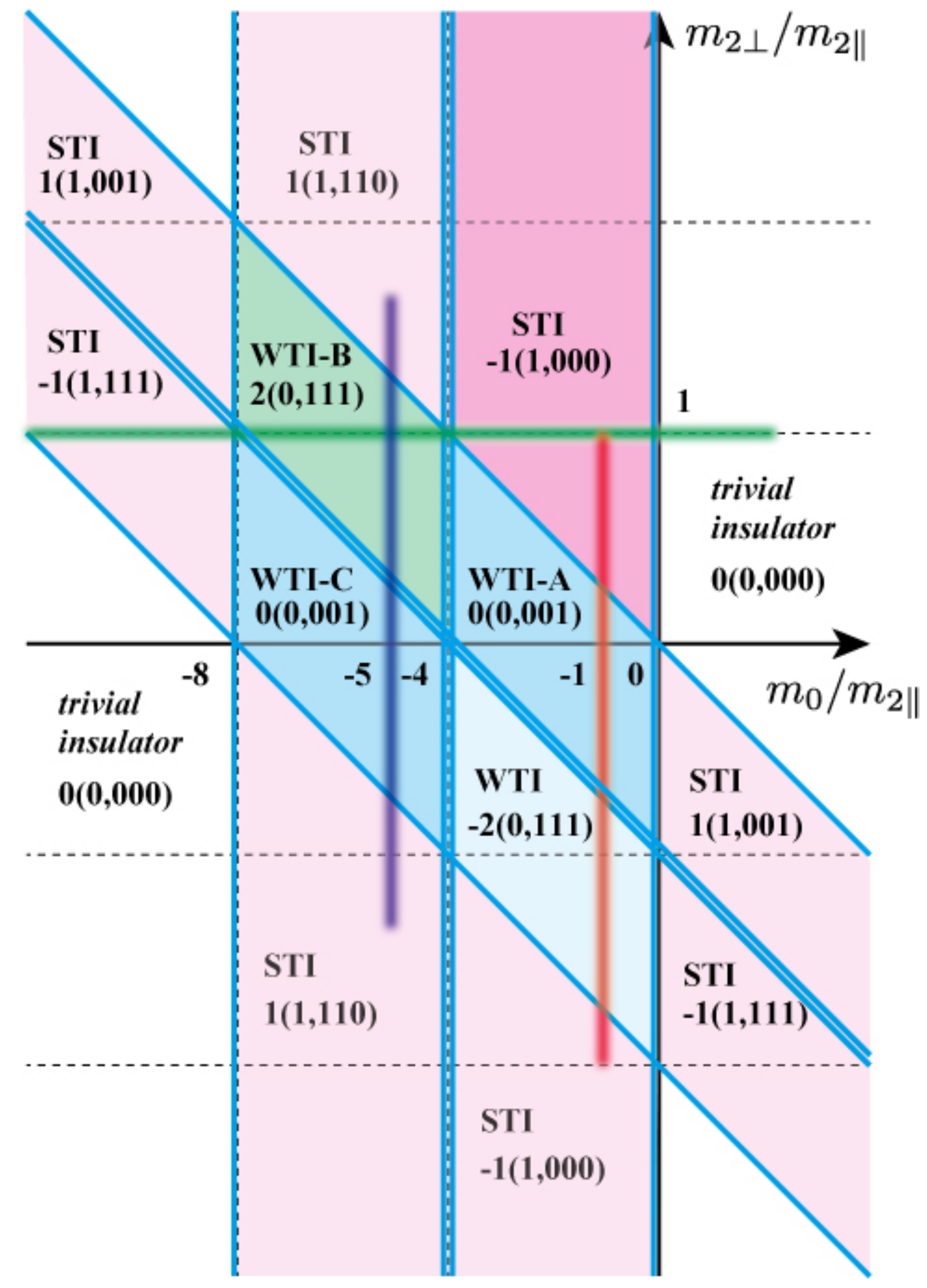}
\end{center}
\caption{The phase diagram
of the Wilson-Dirac type effective tight-binding Hamiltonian
given in Eqs. (\ref{H_bulk}), (\ref{mass_k}).
Notice the anisotropy of our hopping parameters [see Eqs. (\ref{uniaxial})].
In each of the  strong (STI) and weak (WTI) topological insulator phases, 
together with the nature of the specific phase,
the four $\mathbb Z_2$ indices $\nu_j$ ($j=0,1,2,3$)
and the winding number $N_3$ are shown,
as $N_3 (\nu_0, \nu_1 \nu_2 \nu_3)$.
The solid lines representing the phase boundaries 
correspond to closing of the bulk energy gap.}
\label{phase_diagram}
\end{figure}
%%%%%%%%%%%%%%%%%%%%%%%%%%

\section{Model and its phase diagram---engineering the weak indices}

As a concrete realization of strong and weak topological insulators
with specific strong and weak indices, 
$\nu_0$ and $\vec{\nu}=(\nu_1, \nu_2, \nu_3)$,
we consider
as given in Eq. (\ref{H_bulk}),
a Wilson-Dirac type effective Hamiltonian for a 3D topological insulator
implemented on a cubic lattice.\cite{Liu_PRB, Liu_nphys}
Since we will be interested in the analysis of WTI phases
with anisotropic weak indices,
we choose
the mass parameters $m_{2x}$, $m_{2y}$, $m_{2z}$
appearing in the Wilson term [see Eq. (\ref{mass_k})]
to be anisotropic.
%In later analyses, we will assume, for mathematical simplicity, the $\epsilon (\bm k) \bm 1$ term in Eq. (\ref{H_bulk}) to be null.
%This simplification allows us to introduce a $\mathbb Z$-type winding number $N_3$.
%Then, the strong index $\nu_0$ of the system is identified as $N_3 \mod 2$. 
%Independently of this, we calculate the strong and weak $\mathbb Z_2$-indices, using the Fu-Kane's formula \cite{FuKane} to establish the phase diagram depicted in FIG. \ref{phase_diagram}.

\subsection{The Wilson-Dirac type effective Hamiltonian}
Let us consider the following
Wilson-Dirac type effective Hamiltonian for a 3D topological insulator
implemented on a cubic lattice:
\begin{equation}
H_{\rm bulk} = \epsilon (\bm k) \bm 1 +
\tau_x m(\bm k) + \tau_y \sigma_\mu A_\mu \sin k_\mu,
\label{H_bulk}
\end{equation}
where $\epsilon (\bm k)$ is an even function of $\bm k$, and
\begin{equation}
m(\bm k) = m_0 + 2 m_{2\mu} (1-\cos k_\mu).
\label{mass_k}
\end{equation}
In Eqs. (\ref{H_bulk}) and (\ref{mass_k})
a summation over the repeated index $\mu$ ($=x,y,z$)
is not shown explicitly. 
The model specified by this couple of equations
can be regarded as a tight-binding model 
with only the nearest neighbor hopping, 
determining the structure of the energy bands
over the entire 3D Brillouin zone (BZ).
Eq. (\ref{H_bulk}) can be regarded as a $4\times 4$ matrix,
spanned by two types of Pauli matrices $\bm \sigma$ and $\bm \tau$
each representing physically real and orbital spins, respectively.
Compared with a more generic representation of the Dirac Hamiltonian in terms of the
''$\bm \gamma$-matrices'',
we have chosen in Eq. (\ref{H_bulk})
''$\gamma_0$'' coupled to the mass term $m(\bm k)$
associated only with an {\it orbital} spin $\tau_x$.

The mass term (\ref{mass_k}) represents
(a half of) the band gap at time-reversal invariant momenta (TRIM), 
$\bm k = \bm k_0$,
satisfying $-\bm k_0 = \bm k_0 + \bm G$,
with $\bm G$ being a reciprocal lattice vector,
corresponding either to a normal or an inverted gap,
depending on the relative sign of $m_0$
and the coefficient of the quadratic (Wilson) term at a given TRIM.
By investigating this feature of band inversion at the eight TRIM
as varying the mass parameters,
one can identify\cite{FuKane}
various weak and strong TI phases
characterized by
strong and weak indices, $\nu_0$ and $\vec{\nu}=(\nu_1, \nu_2, \nu_3)$.
Phase boundaries between such
topologically distinguishable insulating phases
correspond necessarily to closing of the bulk energy gap.

Known examples of 3D topological insulators
are layered materials, 
exhibiting, in the leading order approximation, uniaxial anisotropy in the crystal $c$-axis.
\cite{BiSe, BiTe, YA_PRL, AK_PRL, YA_PRB}
To reflect this feature in
the effective tight-binding model, i.e., in Eqs. (\ref{H_bulk}), (\ref{mass_k})
we assume that our model parameters have the same
uniaxial anisotropy. \cite{Liu_PRB}
Especially,
the three mass parameters
$m_{2x}$,  $m_{2y}$, $m_{2z}$ are classified to
two types: $m_{2\parallel}$ and $m_{2\perp}$,
depending on whether the corresponding hopping direction
is, either {\it parallel} or {\it perpendicular} to the stacked layers
of the crystal.
Clearly, the correspondence depends on
the relative orientation of the crystal growth axis
and our cartesian coordinates;
e.g., when the crystal $c$-axis is oriented to the direction of
$z$-axis,
\begin{eqnarray}
m_{2\perp} &=& m_{2z},\ \ 
m_{2\parallel} = m_{2x} =  m_{2y},
\nonumber \\
A_\perp &=& A_z,\ \
A_\parallel =A_x =A_y.
\label{uniaxial}
\end{eqnarray}
Independently of this choice of the relative orientation,
our control parameters
for specifying topologically different phases
are relative magnitudes of
$m_0$, $m_{2\perp}$ and $m_{2\parallel}$.
Then,
by studying the feature of band inversion at eight TRIM 
as a function of these control parameters,
\cite{FuKane}
one can deduce
%(see Appendix A for details)
the phase diagram of the model.
FIG. \ref{phase_diagram} shows such a phase diagram depicted in
the ($m_0/m_{2\parallel}, m_{2\perp}/m_{2\parallel}$)-plane.
%Notice that some (anisotropic) weak indices indicated in the figure are due to the specific choice (\ref{uniaxial}).

\subsection{Phase diagram}

FIG. \ref{phase_diagram} shows the phase diagram
of the Wilson-Dirac type effective tight-binding Hamiltonian
given in Eqs. (\ref{H_bulk}), (\ref{mass_k}).
The uniaxial anisotropy 
of the hopping parameters, as given by  Eqs. (\ref{uniaxial}),
is taken into account.
Each of the STI and WTI phases are characterized by
four $\mathbb Z_2$ indices.
The calculated winding number $N_3$ 
%\textcolor{blue}{
(see Appendix A)
is also shown.
Solid lines, separating neighboring topologically distinct phases, 
indicate closing of the bulk energy gap.
Duplicate lines appearing at the phase boundary
correspond to simultaneous formation of
two bulk 3D Dirac cones.
The duplication is due to the uniaxial choice of the hopping parameters.
To see such specific features, let us focus below on a few 
particular examples of the STI and WTI phases.

Let us first concentrate on the isotropic line
$m_{2\perp}/m_{2\parallel} = 1$ 
in the phase diagram
(indicated as a thick green line in FIG. \ref{phase_diagram}).
The change of the winding number $N_3$ on this line is
shown in the first panel of FIG. \ref{winding}.
Notice that on this line
different STI and WTI phases show
only symmetric weak indices.
At the phase boundaries between
STI and WTI phases,
a double and single solid lines cross,
indicating simultaneous closing of three Dirac cones in the bulk.
%As shown in Table II (see Appendix A)
This occurs at $X$, $Y$, $Z$:
$\bm k_X=(\pi,0,0)$, $\bm k_Y=(0,\pi,0)$, $\bm k_Z=(0,0,\pi)$,
three symmetric points (TRIM) in the 3D BZ.

Stopping at $m_0/m_{2\parallel} = -1$,
let us now vary $m_{2\perp}/m_{2\parallel}$,
i.e., introduce anisotropy in the mass parameters.
On the line $m_0/m_{2\parallel} = -1$
(a thick red line in FIG. \ref{phase_diagram}),
the system is in a STI phase with
$\nu_0=1$ and $\vec{\nu}=(0,0,0)$,
when $m_{2\perp}/m_{2\parallel} > 1/4$.
The anisotropy appears in the weak indices below
this critical value,
$m_{2\perp}= -m_0/4$,
corresponding to band crossing occurs at the $Z$-point,
and the system enters a WTI-A phase
with $\nu_0=0$ and $\vec{\nu}=(0,0,1)$
when $m_{2\perp}/m_{2\parallel} < 1/4$.
In later sections we will quantify 
various manifestations of this
quantum phase transition in the finite size effects.
The situation is similar on the line $m_0/m_{2\parallel} = -5$
(a thick blue line in FIG. \ref{phase_diagram}),
above and below the critical point $m_{2\perp}/m_{2\parallel} = 1/4$,
although in this second example
the transition occurs from an isotropic to an anisotropic WTI phase,
each named, respectively, WTI-B and WTI-C phases.

%%%%%%%%%%%%%%%%%%%%%%%%%%%%%%
\begin{table}[htdp]
\begin{center}
\begin{tabular}{lcccc}
\hline\hline
geometry & \ \ & $x$-PBC & $y$-PBC & $z$-PBC \\
\hline
surfaceless && 1 & 1 & 1 \\
slab && 1 & 1 & 0 \\
(rectangular) prism && 0 & 1 & 0 \\
cubic && 0 & 0 & 0 \\
\hline\hline
\end{tabular}
\end{center}
\caption{Definition of the surfaceless, slab, (rectangular) prism and cubic geometries. Here, to avoid confusion in the 
terminology, we define these different types of geometries in terms of the switching on and off of 
the periodic boundary conditions  (PBC) in the $x$-, $y$- and $z$-directions. In the Table,
``1'' and ``0'' signifies that the PBC in the corresponding direction is, respectively, on and off.
In the latter case, PBC is replaced by the fixed boundary condition (FBC).}
\label{geometry}
\end{table}
%%%%%%%%%%%%%%%%%%%%%%%%%%%%%%

\section{Different origins of the finite-size energy gap}

A single Dirac cone on the surface of a STI is 
topologically protected,
\cite{HasanKane}
and also robust against disorder.
\cite{KN, Bardarson}
In reality, TI samples always have a finite thickness between the two surfaces
of opposing sides.
Imagine a slab-shaped sample ({\it c.f.} Table \ref{geometry}), 
which we assume infinitely large,
neglecting 
the existence of side surfaces.
In such a slab geometry,
STI bears a pair of surface Dirac cones,
each localized in the vicinity of the two opposing surfaces.
%\textcolor{yellow}{(by the word, Dirac cone,  we primarily refer to a feature of the spectrum, i.e., of the eigenvalue, which is necessarily accompanied by a corresponding eigenvector, or the wave function that is localized in the vicinity of the surface)}
These two ``Dirac cones'' do not communicate, 
and consequently remain gapless,
as far as the thickness of the slab is much larger than the
penetration of the surface state %(Dirac cones)
into the bulk
%\textcolor{blue}{
[see Appendix B for an extensive discussion on
the penetration of the surface wave function in the slab geometry;
see also Refs. \cite{Shen_PRL, Shen_PRB, Linder, Shen_NJP}].
%We have mentioned in the Introduction that in the slab geometry the surface electronic spectrum suffers only from this type of gap opening due to mixing of the wave functions on the opposing sides.
%This type of finite-size energy gap also decays exponentially as a function of the thickness of the slab.

%One can equally interpret this in a somewhat inverted manner.
%Albeit exhibiting a small finite-size energy gap or a mass, the surface Dirac states appear in the spectrum of a slab-shaped TI, well separated from the rest of the spectrum, inside the window created by the bulk energy gap.
%In this sense, the ``gaplessness'' of the surface state is still protected by the bulk energy gap, as far as its magnitude exceeds significantly that of the size gap.
In a sense
this gaplessness is also protected %by the system's geometry, i.e.,
by the very slab geometry.
In the case of a sample of more realistic shape with typically side surfaces
({\it cf.}, cases of a prism and a cube; see Table \ref{geometry}),
the same protection is no longer valid.
The side surfaces open %a pair of 
{\it a priori} gapless channels 
allowing for communication between the two initial Dirac cones 
on two surfaces of the slab.
Since
this communication through gapless side surfaces
is much stronger than the one through the gapped bulk
({\it cf.} case of the slab geometry),
it leads to opening of a size gap 
qualitatively more relevant
than the latter case.

Of course, the effects of such side surfaces appear in the transport characteritics
only when an electron can really ``see'' the ends of the sample.
In a macroscopic sample in which
%the coherence length of the electronic wave function 
the (single-particle) relaxation length,
determined, {\it e.g.,} by the inelastic scattering length,
does not exceed the size of the system, 
finite-sizes effects,
corresponding to a length scale smaller than the former,
are naturally smeared out.
In the following sections 
we consider nanowire samples that have a nano-meter scale cross section,
with its circumference sufficiently smaller than
the relaxation length.
Here, we concentrate on the cylindrical geometry, 
imposing additionally a rotational (cylindrical) symmetry.
We also assume
that the system is extended to infinity, or 
(by taking only two of four end surfaces into account)
periodic in the remaining direction.
A symptom of the effects we discuss in this section
may be observed experimentally
in a transport measurement analogous to the one in
Ref. \cite{AB_exp}.

\subsection{Spin-to-surface locking on the cylindrical surface}

The protected surface state of a topological insulator is often cited with another
adjective ``helical''.
The word, helical, stems from a specific feature,
often referred to as spin-to-momentum locking,
\cite{helical}
that the helical state exhibits in momentum space.
Here, we highlight another characteristic of the helical surface state, 
the ``spin-to-surface locking'',
which manifests in {\it real} space,
and when the surface is curved.
The electronic spin in a helical state on such a curved surface
is shown to be locked in-plane to the local tangent of the
surface.
\cite{Ashvin_PRL, Mirlin_PRL, JMoore_PRL, k2}

The spin-to-surface locking can be also regarded as a consequence of 
(spin) Berry phase of $\pi$.
In the case of rotationally symmetric (cylindrical) wire, 
the orbital angular momentum along the axis of the wire
is quantized to be {\it half-odd} integers.
This half-odd integral quantization gaps out the spectrum 
of electronic motion along the wire.
The spin-to-surface locking leads, indeed, 
irrespective of the presence of rotational symmetry,
to opening of the Dirac spectrum.

To be explicit let us consider
the continuum limit of Eqs. (\ref{H_bulk}) and (\ref{mass_k}),
or an effective $\bm k \cdot \bm p$ Hamiltonian
at the $\Gamma$-point ($\bm k =\bm 0$),
\begin{equation}
H_{\rm bulk} = \epsilon (\bm p) \bm 1 + \tau_x m(\bm p) + A \tau_y \sigma_\mu p_\mu,
\label{H_Gamma}
\end{equation}
where $m(\bm p) = m_0 + m_2 \bm p^2$.
Here, we focus on the isotropic case: 
$m_{2\mu} = m_2$ and $A_\mu = A$ for $\mu = x, y, z$.
We also assume $\epsilon (\bm p) = 0$, for simplicity.
We then consider
the eigenvalue problem for Eq. (\ref{H_Gamma}), i.e.,
\begin{equation}
H_{\rm bulk} |\psi \rangle\rangle = E |\psi \rangle\rangle,
\label{bulk_eigen}
\end{equation}
in the cylindrical coordinates:
\begin{equation}
\label{theta_def}
r=\sqrt{x^2+y^2},\ \ \
\phi=\arctan {y\over x}.
\end{equation}
Note that
our TI sample occupies the interior of a cylinder of radius $R$. 
%\textcolor{blue}{
As shown in the Appendix C,
any surface solutions 
$|\bm \alpha \rangle\rangle$
of Eq. (\ref{bulk_eigen})
can be expressed as a linear combination of the two basis
solutions,
\begin{eqnarray}
|\bm r + \rangle\rangle_{\rm dv} =  \rho (r)|\tau_z + \rangle |\bm r + \rangle_{\rm dv},
\nonumber \\
|\bm r - \rangle\rangle_{\rm dv} =  \rho (r)|\tau_z - \rangle |\bm r - \rangle_{\rm dv},
\label{r_dv2}
\end{eqnarray}
where $|\tau_z \pm\rangle$ is an eigenstate of $\tau_z$ with the corresponding eigenvalue $\pm 1$
and
\begin{equation}
\label{r_dv}
|\bm r \pm\rangle_{\rm dv} = {1\over \sqrt{2}}
\left[
\begin{array}{c}
e^{-i\phi/2} \\ \pm e^{i\phi/2} 
\end{array}
\right]
\end{equation}
are two real spin eigenstates pointing either
to the centrifugal ($+\bm r$) or to the centripetal ($-\bm r$) direction.
In Eqs. (\ref{r_dv2}),
$\rho (r)$ is the radial part of the surface wave function
localized in the vicinity of the surface of
the cylinder,
given explicitly in Eq. (\ref{rho_n}). 
In Eqs. (\ref{r_dv2}), (\ref{r_dv}) 
the subscript ``dv'' is added to make explicit that these spinors are double-valued.
In terms of
$|\bm r \pm\rangle\rangle_{\rm dv}$, 
the surface solution $|\bm \alpha \rangle\rangle$ reads
\begin{equation}
|\bm \alpha \rangle\rangle =
\alpha_+ (\phi)
|\bm r + \rangle\rangle_{\rm dv}
+
\alpha_- (\phi)
|\bm r - \rangle\rangle_{\rm dv}.
\label{alpha_bulk}
\end{equation}
Here, the explicit form of the coefficients $\alpha_\pm (\phi)$
is determined by solving the eigenvalue problem for
the following surface effective Hamiltonian,
\begin{equation}
H_{\rm surf} = A \left[ -{1\over R} 
\left(-i{\partial \over \partial\phi}\right) 
\sigma_x + p_z \sigma_y \right],
\label{H_surf}
\end{equation}
i.e.,
\begin{equation}
H_{\rm surf} \bm \alpha (\phi) = E \bm \alpha (\phi),
\label{surf_eigen}
\end{equation}
where
\begin{equation}
\bm\alpha (\phi) =
\left[
\begin{array}{l}
\alpha_+ (\phi)\\
\alpha_- (\phi)
\end{array}
\right].
\label{alpha_surf}
\end{equation}
Notice here that
thanks to the rotational symmetry with respect to the axis of the cylinder
the orbital angular momentum $L_z$ is a good quantum number,
which can be simultaneously diagonalized
with $H_{\rm surf}$ and $p_z$.
In the following, we focus on such surface eigenstates of $L_z$,
which can be represented in terms of $\bm \alpha (\phi)$
introduced in Eqs. (\ref{surf_eigen}), (\ref{alpha_surf})
as
\begin{eqnarray}
\bm\alpha (\phi) &=& \bm\alpha_{L_z, p_z} (\phi)
\nonumber \\
=\left[
\begin{array}{l}
\alpha_+ (\phi) \\
\alpha_- (\phi)
\end{array}
\right] &=& e^{i L_z \phi}
\left[
\begin{array}{l}
\alpha_+ (0) \\
\alpha_- (0)
\end{array}
\right].
\label{alpha_Lz}
\end{eqnarray}
$\alpha_\pm (0)$ is specified by the orientation of the
surface crystal momentum specified by $p_z$ and $p_\phi = L_z /R$.
The corresponding eigenenergy $E$ of $H_{\rm surf}$ 
%\textcolor{blue}{
%(and also of $H_{\rm bulk}$, see Appendix C for why this is indeed the case)
is then specified by $p_\phi$ and $p_z$ as
\begin{equation}
E = E (p_\phi, p_z) = \pm A \sqrt{p_\phi^2 + p_z^2}.
\label{spec_cyl}
\end{equation}

The state $|\bm \alpha \rangle\rangle$ thus given,
and specified by the $\bm \alpha (\phi)$ given in Eq. (\ref{alpha_bulk}),
signifies
a simultaneous eigenstate of $H_{\rm bulk}$, $L_z$ and $p_z$,
which may be also represented
$|L_z, p_z \rangle\rangle$.
%As we further clarify below,
Eq. (\ref{alpha_bulk}) implies that
such a state is an equal-weight superposition of 
the centrifugal and the centripetal spin components
given in Eqs. (\ref{r_dv}),
since $|\alpha_+ (0)|=|\alpha_- (0)|$.
This signifies that when an electron is on the surface of the cylinder
at an angle $\phi$ in the configuration space,
its spin state is constrained onto the local tangent of the cylinder
at this position
(spin-to-surface locking).
While an electron travels around the cylinder
in the configuration space,
the corresponding spin frame also completes a $2\pi$ rotation in the spin space.

\subsection{Half-integral quantization of the orbital angular momentum
and the resulting finite-size energy gap}

Let us reconsider the statue of the angle $\phi$ in different steps of the
formulation.
In the original bulk effective Hamiltonian (\ref{H_Gamma})
the angle $\phi$ purely specifies the position of an electron
in the configuration space.
This is also the case in its eigenstate
$|\bm \alpha \rangle\rangle$.
Therefore, $|\bm \alpha \rangle\rangle$ must be single-valued with respect to the
$2\pi$-rotation of $\phi$,
\begin{equation}
|\bm \alpha \rangle\rangle 
|_{\phi\rightarrow \phi + 2\pi} = |\bm \alpha \rangle\rangle. 
\end{equation}
On contrary, $\phi$ in
$|\bm r \pm\rangle_{\rm dv}$ specifies the direction of real $\mathbb{SU}(2)$ spin.
Therefore, $|\bm r \pm\rangle\rangle_{\rm dv}$ is double-valued with respect to the
$2\pi$-rotation of $\phi$,
\begin{equation}
|\bm r \pm\rangle\rangle_{\rm dv} |_{\phi\rightarrow \phi + 2\pi} = - |\bm r \pm\rangle\rangle_{\rm dv}.
\end{equation}
In Eq. (\ref{alpha_bulk})
these two boundary conditions are compatible, only if
\begin{equation}
\bm \alpha (\phi +2\pi) = - \bm \alpha (\phi),
\label{apbc}
\end{equation}
i.e., the coefficients $\alpha_\pm (\phi)$ are also anti-periodic.
In the light of Eq. (\ref{alpha_Lz}),
this requires,
\begin{equation}
L_z = \pm {1\over 2}, \pm {3\over 2}, \cdots,
\label{Lz_half}
\end{equation}
i.e.,
the {\it orbital} angular momentum $L_z$ is quantized to be
half-odd integers.

Notice also that the double-valuedness of $|\bm r \pm\rangle\rangle_{\rm dv}$
is not essential for the half-integral quantization of $L_z$.
One can equally employ
the single-valued version of Eq. (\ref{r_dv}),
\begin{equation}
|\bm r \pm \rangle_{\rm sv} = {1\over \sqrt{2}}
\left[
\begin{array}{c}
1 \\ \pm e^{i\phi} 
\end{array}
\right],
\label{r_sv}
\end{equation}
which is related to $|\bm r \pm\rangle\rangle_{\rm dv}$ by a simple phase factor,
\begin{equation}
|\bm r \pm \rangle_{\rm sv} = e^{i\phi /2} |\bm r \pm \rangle_{\rm dv}.
\end{equation}
In this single-valued basis the surface effective Hamiltonian acquires an additional
phase factor $\pi$, the spin Berry phase, as
\begin{equation}
\widetilde{H}_{\rm surf} = A \left[ -{1\over R} 
\left(-i{\partial \over \partial\phi}+{1\over 2}\right) 
\sigma_x + p_z \sigma_y \right].
\label{h_sv}
\end{equation}
Then, if one employs the same representation (\ref{alpha_Lz})
for the coefficients $\bm \alpha$,
$L_z$ takes formally integral values,
$L_z = 0, \pm 1, \pm 2, \cdots$.
The corresponding eigenenergy $E = E (p_\phi, p_z)$ can be also
written formally in the same way as in Eq. (\ref{spec_cyl}).
But in that case, $p_\phi$ in the same formula must be reinterpreted as
\begin{equation}
p_\phi = {L_z + 1/2 \over R}.
\label{p_phi}
\end{equation}

We have so far seen that
whether one employs the double-valued [Eq. (\ref{r_dv})]
or the single-valued [Eq. (\ref{r_sv})]
basis,
one finds, as expected, the same
gapped spectrum given by
Eq. (\ref{spec_cyl}) with either
i) $p_\phi = L_z /R$ with half-odd $L_z$ [Eq. (\ref{Lz_half})], or
ii) $p_\phi$ given as in Eq. (\ref{p_phi}) with $L_z = 0, \pm 1, \pm 2, \cdots$.
The magnitude of the energy gap is given by
twice of
\begin{equation}
E_0 = E \left( {1\over 2R}, 0\right) = {A\over 2R} \propto R^{-1}.
\label{gap_cyl}
\end{equation}
This energy gap due to spin-to-surface locking, 
or eventually
to the doubling of the original two Dirac cones through ``side surfaces'' of the cylinder,
decays only {\it algebraically} as a function of
(inversely proportionally to) the circumference of the cylinder.
This {\it enhanced} finite-size energy gap is in marked contrast 
with that of the slab due to mixing of the two surface wave functions
sitting mainly on the opposing sides of the slab and
separated by the bulk energy gap.

%%%%%%%%%%%%%%%%%%%%%%%%%%%%%%
\begin{table*}[htdp]
\begin{center}
\begin{tabular}{lllll}
\hline\hline
cases & type of the phase & parity of $N_z $ & size gap; $N_z $ dependence & gap opening mechanism \\
\hline
(a) & WTI & even & $N_z^{-1}$ & (iii) doubling of Dirac cones due to confinement \\
\hline
(b) & WTI & odd & ``0'' (exponentially small) & (i) mixing of the opposing sides through gapped bulk \\
\hline
(c) & STI & irrelevant & $(N_z + N_x)^{-1}$ & (ii) spin-to-surface locking \\
\hline\hline
\end{tabular}
\end{center}
\caption{Three typical behaviors of the finite-size energy gap in the rectangular-prism shaped samples.}
\label{abc}
\end{table*}
%%%%%%%%%%%%%%%%%%%%%%%%%%%%%%

%%%%%%%%%%%%%%%%%%%%%%%%%%
\begin{figure}
\begin{center}
\includegraphics[width=8cm]{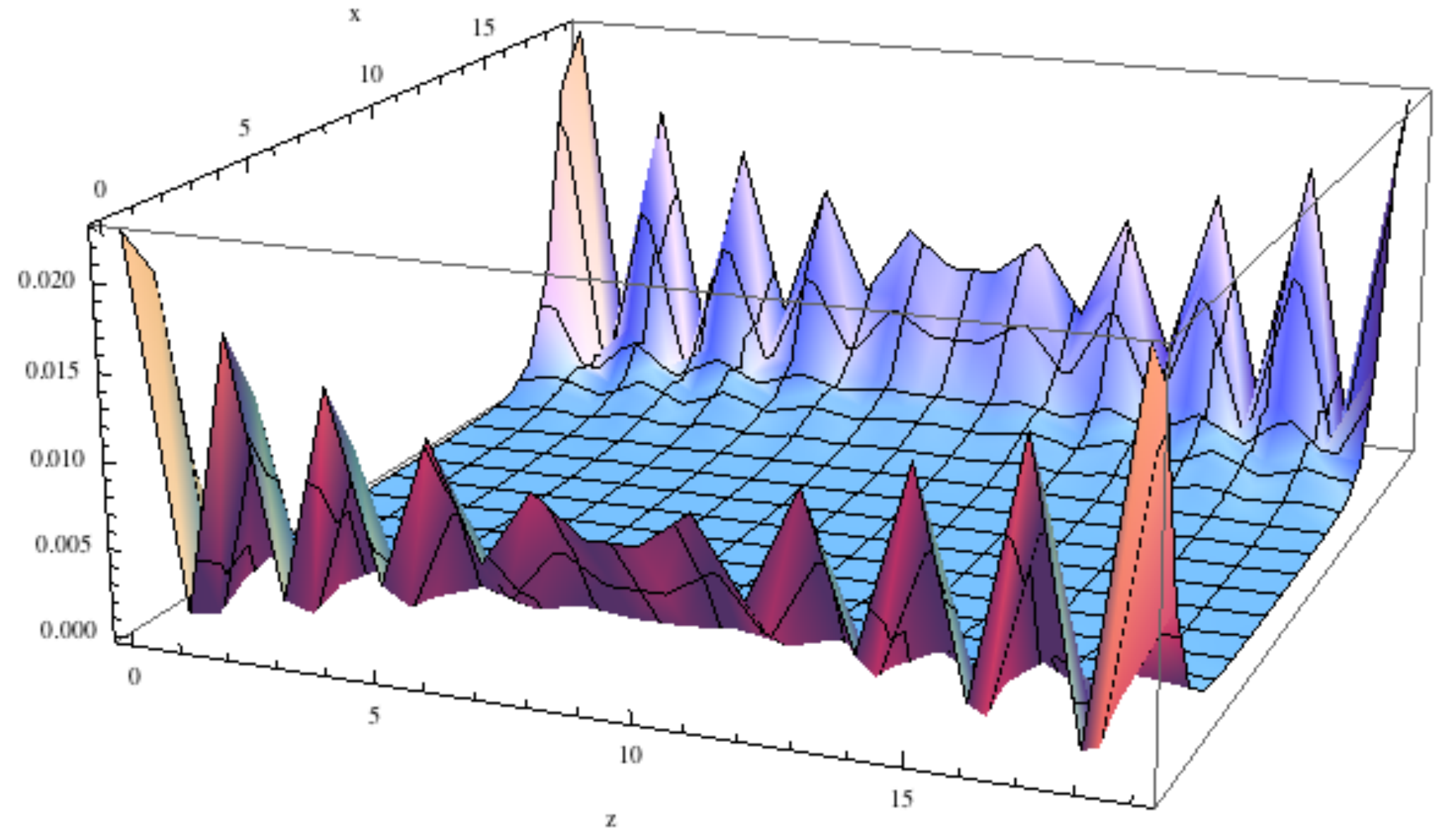}
\includegraphics[width=8cm]{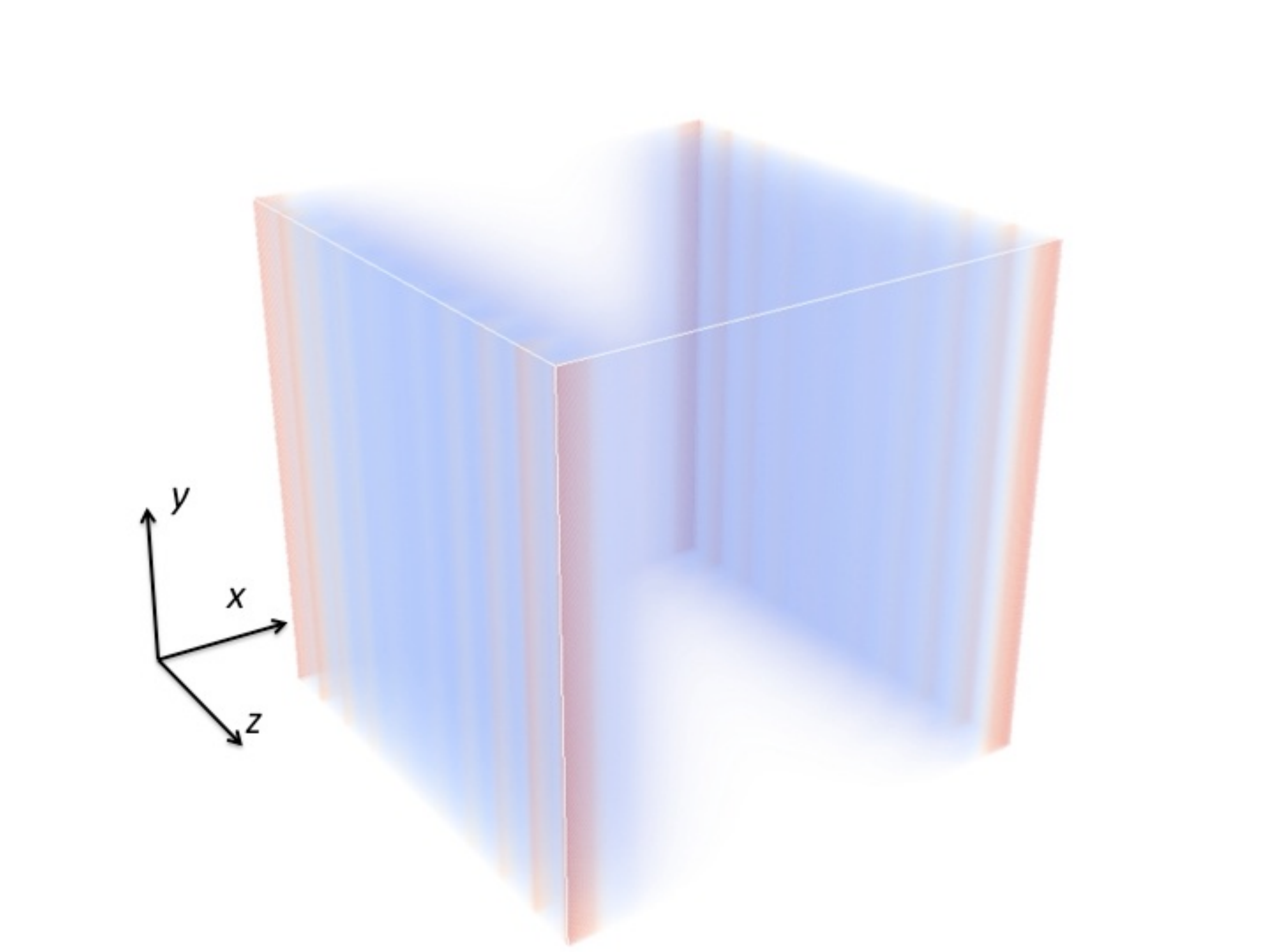}
\end{center}
\caption{Surface wave function
in the rectangular prism geometry
[Eq. (\ref{prism})];
WTI phase ($m_{2z}/m_{2\parallel} = 0.2$) with $N_z$ even.
Upper: the square of the wave function
$|\psi (z,x)|^2$ with $N_z =20$, $N_x =20$ and $k_y = 0$
is plotted in the $(z,x)$-plane.
Spin and orbital indices are summed over.
$A_\perp = A_\parallel =1$.
Lower: $|\psi (x,y,z)|^2$ is plotted in the 3D $(x,y,z)$-space.
The front, upper and right surfaces correspond, respectively, 
to the ones normal to $(-1,0,0)$, $(0,1,0)$ and $(0,0,1)$.
Fixed boundary condition (FBC) in the $z$- and $x$-directions.
PBC in the $y$-direction.
%The choice of other parameters are specified in the main text.
}
\label{wf0100_even}
\end{figure}
%%%%%%%%%%%%%%%%%%%%%%%%%%

%%%%%%%%%%%%%%%%%%%%%%%%%%
\begin{figure}
\begin{center}
\includegraphics[width=7cm]{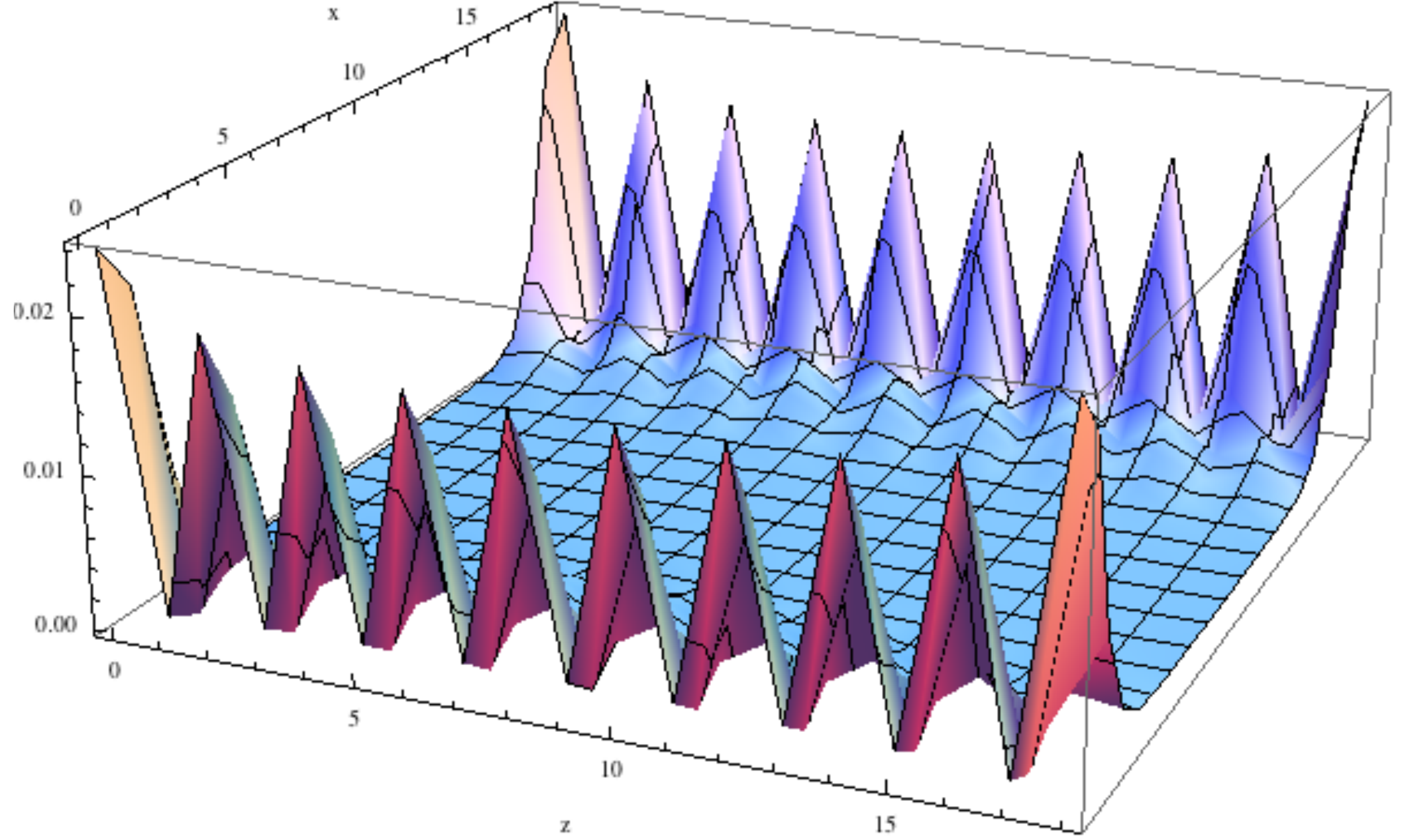}
\includegraphics[width=8cm]{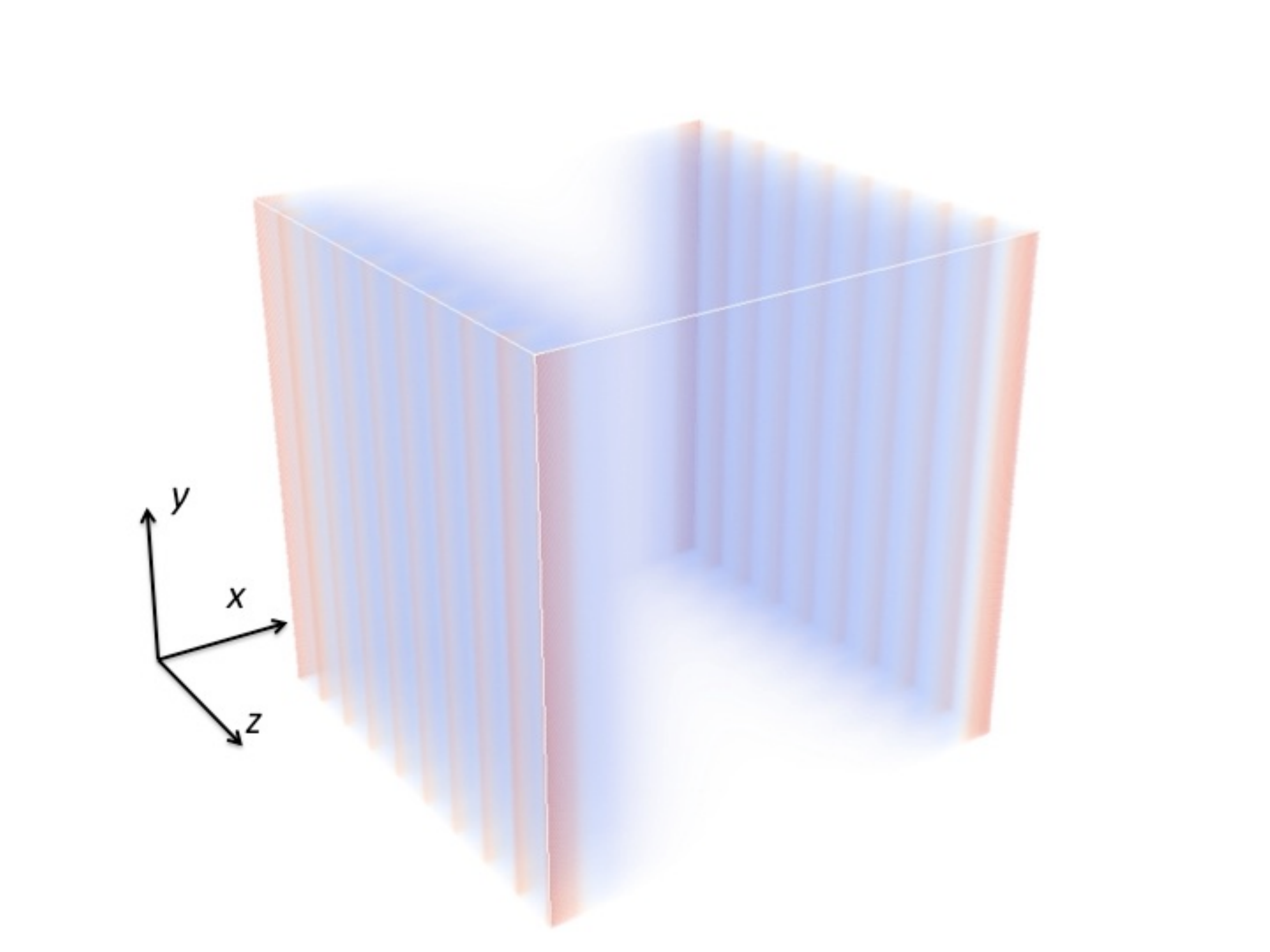}
\end{center}
\caption{Plots of the surface wave function in the rectangular prism geometry
analogous to FIG. \ref{wf0100_even}.
Case of $N_z$ odd ($N_z =19$). WTI phase.
$N_x =20$, $k_y = 0$, $A_\perp = A_\parallel =1$.
FBC in the $z$- and $x$-directions.
PBC in the $y$-direction.}
\label{wf0100_odd}
\end{figure}
%%%%%%%%%%%%%%%%%%%%%%%%%%

%%%%%%%%%%%%%%%%%%%%%%%%%%
\begin{figure}
\begin{center}
\includegraphics[width=8cm]{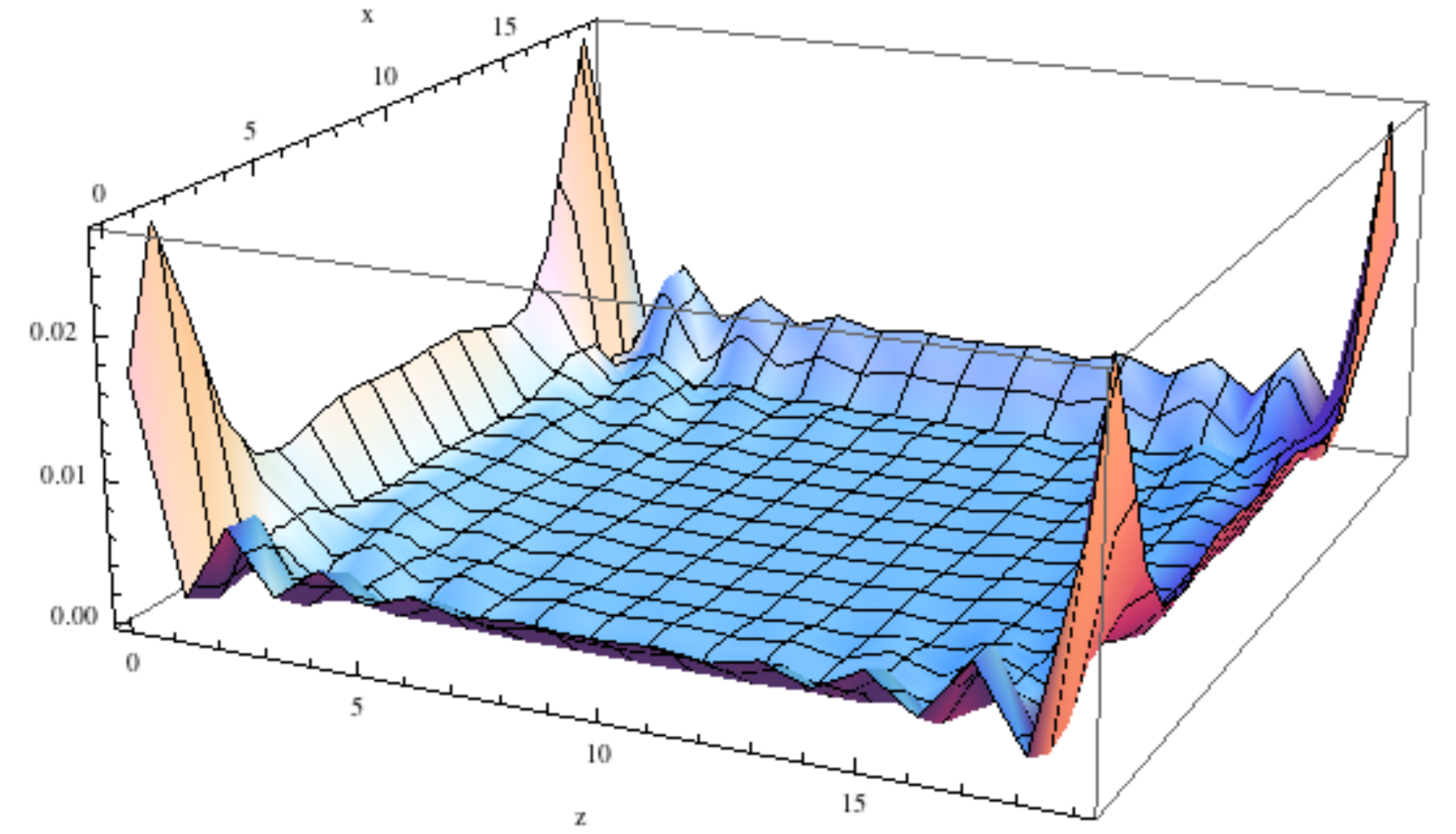}
\includegraphics[width=8cm]{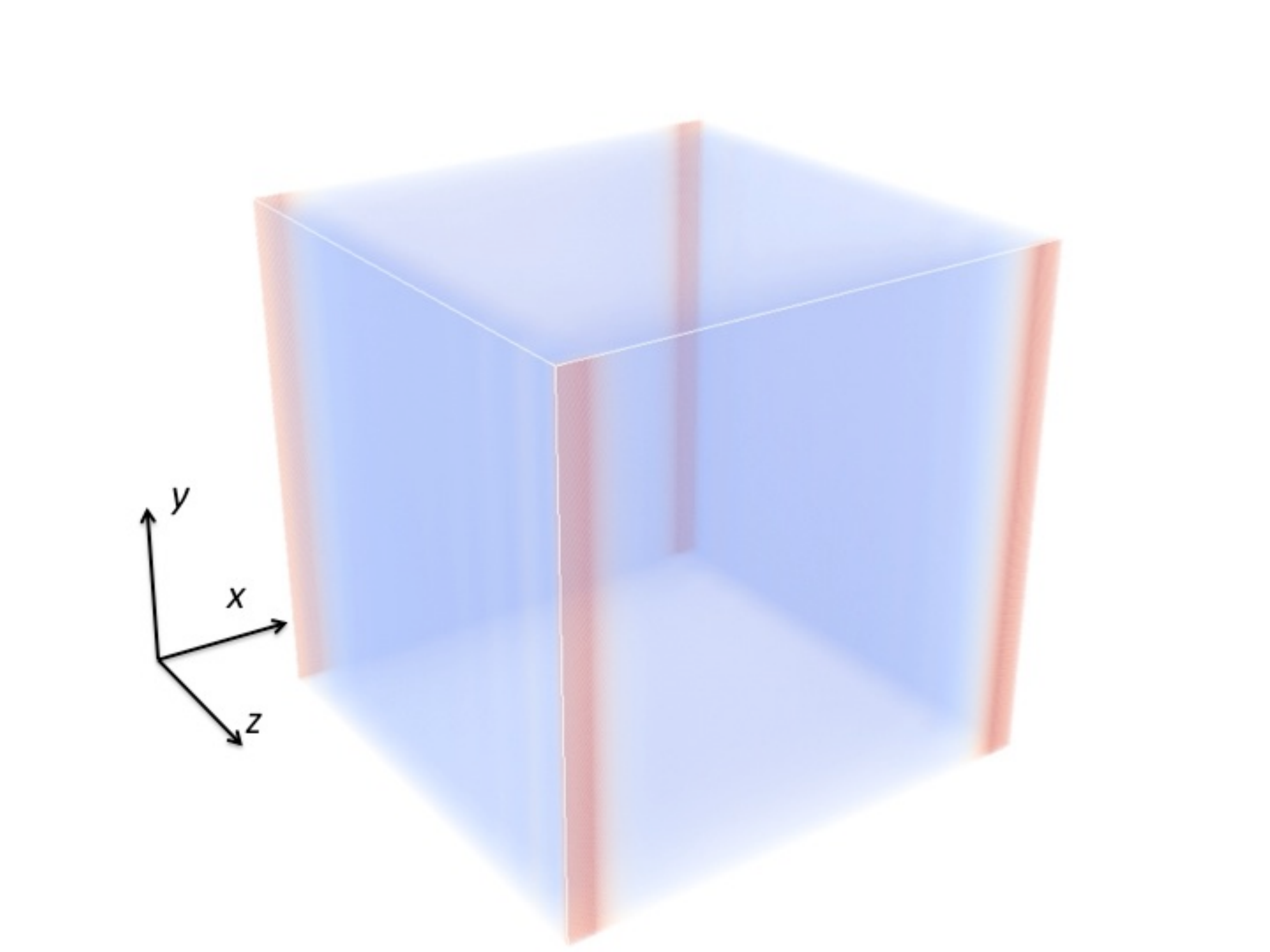}
\end{center}
\caption{Plots of the surface wave function in the STI case ($m_{2z}/m_{2\parallel} = 0.3$);
plots similar to FIG. \ref{wf0100_even} and FIG. \ref{wf0100_odd}.
Here, the surface wave function is extended over
all the four facets of the prism.}
\label{wf1000}
\end{figure}
%%%%%%%%%%%%%%%%%%%%%%%%%%

\section{Case of the rectangular prism geometry}

In the previous section, 
we have considered an idealized case of the cylindrical geometry,
to demonstrate how spin-to-surface locking leads to opening of the finite-size energy gap.
With the rotational (cylindrical) symmetry hypothesized,
the cylindrical geometry was best suited for
analytic considerations of the surface state.
Here, we attempt to realize an equivalent situation in numerical experiments
in terms of the tight-binding simulation.
For that purpose, we consider rather prism-shaped samples
whose cross section on the plane normal to the axis of the (right) prism
is a rectangle rather than a circle. %(see FIG. \ref{prism_geo}).
From the viewpoint of topology, such a rectangular prism shape is
a natural implementation \cite{k2}
of the cylinder-like geometry
on the cubic lattice.

In addition to that aspect as a substitute of a cylinder,
there is also a more positive reason we focus here
on this rectangular prism geometry.
In the last few sections,
throughout the comparison of slab and cylinder,
we have seen that
preventing the communication of two Dirac cones sitting on the opposing sides
of the sample
helps protecting the gaplessness of Dirac cones.
%especially by eliminating the gapless channel in the side surfaces.
We have so far discussed
such switching on and off of this communication channel
by changing the system's (global) geometry.
Here, in this section,
a new element comes into play, the weak indices.
As mentioned in the Intoruduction,
the weak indices have the potentiality of
excluding a gapless Dirac cone from a surface 
oriented in a particular direction,
i.e., that of the weak vector,
$\vec{\nu} = (\nu_1, \nu_2, \nu_3)$.

Folded surfaces of the rectangular prism geometry are
more adapted
for implementing a weak vector
as a means for eradicating the ``dangerous'' gapless channels
from the targeted side surfaces.
Another characteristic of the WTI surface state is that it exhibits
even number of Dirac cones.
These two features combine to make
gaplessness of the surface state of a prism-shaped WTI 
a rather subtle issue, which depends intricately 
on the geometry and on the nature of weak indices.
Depending on the relative orientation between the weak vector and
the surfaces of the rectangular prism and on the size of the prism,
non-compatibility of the surface wave function 
with a specific boundary condition imposed by the geometry
leads to, or not to opening of a finite-size energy gap.

The system we consider here has a shape of rectangular prism
extended in the $\hat{\bm y}$-direction. %(see FIG. \ref{prism_geo}).
We assume that the prism is infinitely long, or periodic,
without end surfaces.
Each cross section of the system at fixed $y$ 
is restricted to a rectangular area
of size $N_z\times N_x$ in the $(z,x)$-plane:
\begin{equation}
1\le z \le N_z,\ 1\le x \le N_x. 
\label{prism}
\end{equation}
The system has two surfaces ($\hat{\bm x}$-surfaces) at $x=1$ and $x=N_x$ 
normal to $\hat{\bm x}= (1,0,0)$
and two others ($\hat{\bm z}$-surfaces) 
at $z=1$ and $z=N_z$ normal to $\hat{\bm z}= (0,0,1)$.
We assume translational symmetry in the $\hat{\bm y}$-direction;
$k_y$ is a good quantum number.
As for the anisotropy of bulk topological insulators,
we consider the case of mass parameters with uniaxial-type anisotropy
as given in Eq. (\ref{uniaxial}).
In the WTI phase, this corresponds to the case of stacked 2D TI layers
piled up in the $z$-direction.

In the following, we will mainly focus on the WTI phase with a specific weak vector
$\vec{\nu}= (0,0,1)$ normal to the $\hat{\bm z}$-surfaces.
Then,
gapless Dirac cones are completely eliminated from these surfaces,
at least in the limit of infinitely large surfaces.
In the prism geometry (\ref{prism}), 
the wave function of the corresponding surface state 
has a finite amplitude only on $\hat{\bm x}$-surfaces, 
barely penetrates into the $\hat{\bm z}$-side.
The Dirac cones forced to be localized in each of the
$\hat{\bm x}$-surfaces are subject to a particular boundary condition
imposed by this combination of the prism geometry and the weak vector.
Compatibility or non-compatibility of the surface wave function
with this specific boundary condition
leads to an even/odd feature with respect to $N_z$ 
(width of the $\hat{\bm x}$-surfaces)
of the finite-size energy gap in the WTI phase.
After reviewing three typical situations we encounter
in the analysis of the size gap in the  WTI and STI phases,
we describe the nature of even/odd feature
in the spirit of $k\cdot p$ approximation.

%%%%%%%%%%%%%%%%%%%%%%%%%%
\begin{figure}
\begin{center}
\includegraphics[width=7cm]{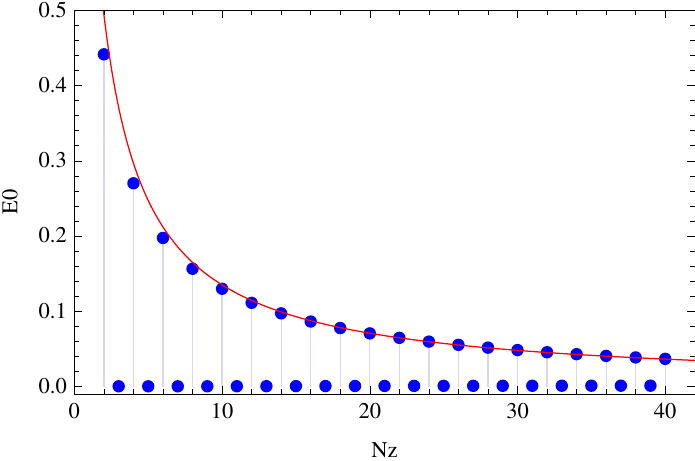}
\includegraphics[width=7cm]{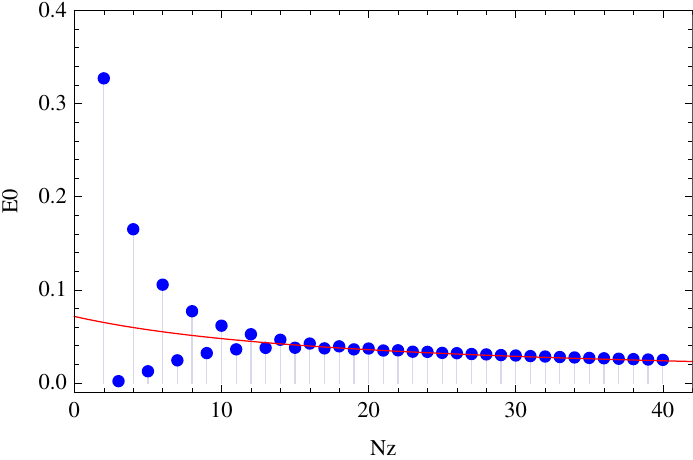}
\end{center}
\caption{Even/odd feature in the finite-size energy gap (case of WTI).
The mass parameters are
on the (red) line $m_0/m_{2\perp}=-1$ of the phase diagram (FIG. \ref{phase_diagram}),
slightly below (WTI-A case, upper panel) 
and above (STI case, lower panel) 
the phase boundary at $m_{2\parallel}/m_{2\perp}=1/4$.
The gap is plotted as a function $N_z$.
In the WTI-A case:
$m_{2z}/m_{2\parallel} = 0.2$,
$E_0 = E_0 (N_z)$ shows an even/odd feature, and
for $N_z$ even the gap scales as $\sim (N_z +1)^{-1}$.
In the STI case:
$m_{2\perp}/m_{2\parallel} = 0.3$,
a weak even/odd feature for small $N_z$ is washed out as $N_z$ increases,
and the gap scales %approximately 
as $\sim (N_z+N_x)^{-1}$. $N_x = 20$. $A_\perp = A_\parallel =1$.}
\label{gap_Nz}
\end{figure}
%%%%%%%%%%%%%%%%%%%%%%%%%%

\subsection{Even/odd feature in the WTI phase}

The three typical situations we investigate 
are the cases of
\begin{itemize}
\item 
WTI with $N_z$ even [case (a)],
\item
WTI with $N_x$ odd [case (b)], and
\item
STI [case (c)].
\end{itemize}
The three cases are also listed in Table \ref{abc}.
In our model, Eqs. (\ref{H_bulk}),  (\ref{mass_k}), 
and in the geometry employed,
the three situations can be realized by a small change of parameters.
As for the concrete choice of parameters,
we use here the following double standard.
\cite{Prodan}
We first use the ``theoretical values'' that varies on the lines
indicated in FIG. \ref{phase_diagram}
for the demonstration of crossover from type (c) to type (a),
and from type (c) to type (b) behaviors.
We believe that use of these theoretical values help
understanding the nature of the phenomenon in the light of the phase diagram.
Then, in the actual computation of the size gap, we also use
''experimental values'' of the parameters that are
deduced from experimental data for Bi$_2$Se$_3$.\cite{Liu_PRB, Ebi}

The three situations can be easily contrasted by
the shape of the surface wave function.
In the WTI phase (FIG. \ref{wf0100_even} and FIG. \ref{wf0100_odd})
the amplitude of the surface wave function concentrates on 
the two $\hat{\bm x}$ surfaces.
The weak vector $\vec{\nu}$ is here pointed in the direction $\hat{\bm z}$,
expels the surface state from the sides normal to $\hat{\bm z}$.
In the STI phase (FIG. \ref{wf1000}), on contrary,
the surface state is extended over all the four surfaces.
In these figures
the square of the total amplitude of the surface wave function,
\begin{equation}
|\psi (z,x)|^2 = \sum_{j=1}^4 |\psi_j (z,x)|^2,
\end{equation}
is plotted at each point on a cross section
(the system is translationally invariant in the $y$-direction).

Let us focus on more detailed structures of
the shape of the surface wave function in the WTI phase,
and compare the cases of $N_z$ even (FIG. \ref{wf0100_even})
and $N_z$ odd (FIG. \ref{wf0100_odd}).
On the two $\hat{\bm x}$ surfaces,
the wave function shows a regular pattern,
vanishing practically at every other layer,
when $N_z$ is odd,
whereas in FIG. \ref{wf0100_even}
it is concave shaped (case of $N_z$ even).

This even-odd feature appears more clearly in the behavior of
the finite-size energy gap
(see FIG. \ref{gap_Nz})
On the (red) line $m_0/m_{2\parallel}= -1$ 
of the phase diagram (FIG. \ref{phase_diagram}),
slightly below [WTI-A: (0,100)] 
and above [STI: (1,000)] 
the phase boundary at $m_{2\perp}/m_0= -1/4$
the gap is plotted as a function $N_z$ (the number of stacking layers).
In the WTI case:
$m_{2z}/m_{2\parallel} = 0.2$,
$E_0 = E_0 (N_z)$ shows an even/odd feature, and
for $N_z$ even the gap scales as $\sim (N_z +1)^{-1}$.
In the STI case:
$m_{2z}/m_{2\parallel} = 0.3$,
a weak even/odd feature for small $N_z$ is washed out as $N_z$ increases,
and the gap scales as $\sim (N_z+N_x)^{-1}$.
In a sense,
depending on the parity of the number of stacked layers,
the system becomes either trivial (gapped, when $N_z$ even)
or gapless (when $N_z$ odd).
Physically, this even/odd feature stems from the fact that
WTI can be viewed as stacked layers of 2D quantum spin Hall states
(here, stacked in the $z$-direction).

%%%%%%%%%%%%%%%%%%%%%%%%%%
\begin{figure}
\begin{center}
\includegraphics[width=7cm]{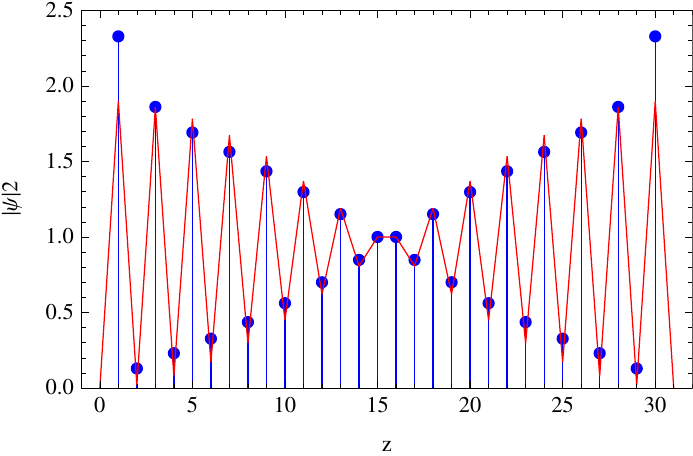}
\end{center}
\caption{Shape of the surface wave function: tight-binding model
vs. $k\cdot p$-approximation. $|\psi (z)|^2$, the squared amplitude 
of the surface state wave function at $x=1$
(and in the case of $k_z =0$) 
is plotted for the case of $N_z$ even ($N_z=30$, blue points).
A continuous red curve is the prediction of $k\cdot p$ theory [cf. Eq. (\ref{psi^2_x})].
As in FIG. \ref{wf0100_even}, the mass parameters are chosen to be
$m_0 /m_{2\parallel} = -1$, $m_{2\perp}/m_{2\parallel} = 0.2$;
other parameters are also set as in FIG. \ref{wf0100_even}.}
\label{wf0100_kp}
\end{figure}
%%%%%%%%%%%%%%%%%%%%%%%%%%

\subsection{Effective surface $\bm k\cdot\bm p$ theory}

A single Dirac cone cannot be confined
({\it cf}. Klein tunneling).
This applies to the STI phase we have considered in Sec. III,
in which any surface state, instead of being terminated at the end of a plane,
continues to the adjacent ones, covering the entire surface.
In the WTI phase,
typically two Dirac cones appear on its surfaces,
{\it i.e.}, there are ``valleys.''
In that case,
one can confine them in a finite area of the surface.
Let us sketch explicitly how this is possible.

A typical situation we focus on below is the case in which
two side faces of the prism is normal to the weak vector $\vec{\nu}$,
implying that there is no Dirac cone on these surfaces.
%(the spectrum is {\it a priori} gapped).
In such a situation,
the wave function of the WTI surface state
has a finite amplitude only on the remaining two surfaces 
parallel to $\vec{\nu}$,
%(here, we neglect the presence of end surfaces),
barely penetrates into the side normal to $\vec{\nu}$.
The key observation here is that the latter can be 
regarded as a ``boundary condition''
for the wave function that lives mainly
on the primary parallel surfaces.

Let us consider a simple and concrete example.
In the WTI-A phase, shown in FIG. \ref{phase_diagram},
only two $\hat{\bm x}$-surfaces 
are compatible with the presence of gapless Dirac cones;
the remaining $\hat{\bm z}$-surfaces are normal to $\vec{\nu}= (0,0,1)$.
We consider the reciprocal space of a $\hat{\bm x}$-surface,
spanned by $k_y$ and $k_z$;
here, we tentatively disregard the presence of $\hat{\bm z}$-surfaces,
pretending as if the translational symmetry in the $z$-direction
is still present.
Then, on this $\bm k = (k_y, k_z)$-plane,
two Dirac points appear in the spectrum
at $\bm k_1 = (0,0)$ and at $\bm k_2 = (0,\pi)$.
The spectrum of the rectangular prism
is obtained, in a crude approximation,
by projecting $E = E (k_y, k_z)$ 
in the $(k_y, k_z)$-plane onto the $k_y$-axis.
When two Dirac cones are superposed in this projection,
a more careful treatment on the boundary condition
at the corner to the $\hat{\bm z}$-surfaces is needed (see below).

The $\hat{\bm x}$-plane on which we focus
is bounded by the $\hat{\bm z}$-surfaces.
Penetration of a surface state into the $\hat{\bm z}$-sides
is incompatible with the weak vector, $\vec{\nu}= (0,0,1)$.
This may be described by a boundary condition on
the surface wave function $\psi (y,z)$
on the $\hat{\bm z}$-side,
\begin{equation}
\psi (y,z=0) =0,\ \
\psi (y,z=N_z +1) =0.
\label{bc_z}
\end{equation}
In the $\bm k\cdot\bm p$  approximation, the wave function $\psi(y,z)$ 
can be constructed by superposing contributions from one valley 
surrounding a Dirac point at $\bm k_{1}$ 
and from another located
at $\bm k_{2}$.
As our system is translationally invariant in the $y$-direction, 
$\psi (y,z)$ is
expressed in the form of 
\begin{equation}
\psi(y,z) = e^{i k_{y}y}\chi(z),
\label{psi_chi}
\end{equation}
where $\chi (z)$ should be chosen
to satisfy the boundary conditions (\ref{bc_z}).
This is allowed only when 
the $y$-components of 
$\bm k_1$ and of $\bm k_{2}$ are identical as
$\bm k_1 = (k_0, k_1)$ and $\bm k_2 = (k_0, k_2)$.
This is indeed the case in the WTI-A phase,
where $k_0 =0$, $k_1 =0$ and $k_2 =\pi$.
The superposition yields 
\begin{equation}
\chi(z) = e^{i(k_{1}+p_{1})z}-e^{i(k_{2}+p_{2})z},
\label{chi}
\end{equation}
where $p_{1}$ and $p_{2}$ are small displacements 
from the corresponding Dirac points.
Note that this automatically satisfies the boundary condition at $z=0$.
If $\chi (z)$ with $p_{1}=p_{2}=0$ 
(i.e., the superposition of the wave functions just at the two Dirac points) 
is compatible with the other boundary condition
at $z=N_{z}+1$, 
the resulting wave function has the zero energy eigenvalue at $k_{y}=k_{0}$,
resulting in the gapless surface states.
This occurs typically at $N_z$ {\it odd}, and
in the WTI-A phase with $k_1 =0$ and $k_2 =\pi$.
Contrastingly, if finite displacements (i.e., $p_{1},p_{2} \neq 0$) are necessary
to satisfy the boundary condition, a finite size gap inevitably appears.
Naturally, the latter applies to the case of $N_z$ {\it even}.
These two contrasting behaviors explain the nature of
the even/odd feature demonstrated in FIG. \ref{gap_Nz}.

Let us further quantify the case of $N_z$ even.
To fulfill the requirement of Eq. (\ref{bc_z}) we set $p_1= -p_2=q$.
The boundary condition at $z=N_z +1$ is satisfied, if
\begin{equation}
q = \pm {n\over 2(N_z +1)}\pi,
\label{q}
\end{equation}
and $n$ being an odd integer.
The lowest energy solution with $n=1$ determines the energy gap
to be,
\begin{equation}
E_0 = {A\over 2(N_z +1)}\pi,
\label{gap_kp}
\end{equation}
i.e., $E_0$ scales as $(N_z +1)^{-1}$ for $N_z$ even
within the range of validity of the $\bm k\cdot\bm p$-approximation.
Eq. (\ref{gap_kp}) allows for comparing 
the above simple effective theory with the calculated spectrum.
This is done in FIG. \ref{gap_Nz}
by plotting the energy gap obtained by numerical diagonalization of the
corresponding tight-binding model
against the postulated scaling of Eq. (\ref{gap_kp}). 

A similar comparison can be made for the shape of the surface wave function.
Plugging Eq. (\ref{q}) with $n=1$ back into Eq. (\ref{chi}) one finds,
\begin{equation}
|\chi (z)|^2 = 4 \sin^2 \left[{N_z \pi \over 2(N_z +1)}z\right].
\label{psi^2_x}
\end{equation}
The shape of this envelop function is to be compared with
the calculated value of the amplitude of the surface state
eigenspinor at $x=1$,
which is shown in FIG. \ref{wf0100_kp}.

It is suggestive to apply the above $\bm k\cdot\bm p$ effective theory
to the case of WTI-B and WTI-C phases.
(see FIG. \ref{phase_diagram}).
These two topologically different WTI phases
appear on the blue line $m_0/m_{2\parallel}=-5$
in the phase diagram
with the phase boundary at $m_{2\perp}/m_{2\parallel}= -1/4$.
The crossover of the finite-size energy gap
at the transition between these two WTI phases
is precisely in parallel with the one
between STI and WTI-A phases
(on the red line: $m_0/m_{2\parallel}=-1$ in FIG. \ref{phase_diagram})
we have considered so far.
In the case of WTI-B and WTI-C phases,
The constituent surface Dirac cones on the $\bm k = (k_y, k_z)$-plane
appear at
$\bm k_1 = (\pi,0)$ and at $\bm k_2 = (0,\pi)$ in the WTI-B phase,
and at 
$\bm k_1 = (\pi,0)$ and at $\bm k_2 = (\pi,\pi)$ in the WTI-C phase.
Here, the relative position of the two Dirac cone is essential.
In the case of WTI-C phase,
one can construct the surface wave function (\ref{chi})
compatible with the specific boundary condition (\ref{bc_z}) 
precisely in parallel with the previous case of the WTI-A phase,
simply by replacing $k_0 =0$ with $k_0 = \pi$,
leading to the same even/odd feature.
Notice that the surface Dirac cone in the WTI-C phase
appears in the spectrum of prism geometry $E = E_{\rm prism} (k_y)$
at $k_y =\pi$.

In the case of WTI-B phase,
the two Dirac cones at
$\bm k_1 = (\pi,0)$ and at $\bm k_2 = (0,\pi)$
are projected onto a different point
on the $k_y$ axis,
making the previous construction [Eqs. (\ref{psi_chi}), (\ref{chi})]
impossible.
%{\it i.e.}, impossible to cope with the boundary condition (\ref{bc_z}).
%They do not conspire to satisfy the boundary condition (\ref{bc_z}).
This is, of course, consistent with the fact 
that in the WTI-B phase the surface states are 
not confined to the $\hat{\bm x}$-surfaces.
This observation, in turn, %implies 
reveals that
the relative orientation of the two (even number of)
Dirac cones in the WTI is %determined
indeed imposed 
by the weak indices.
On surfaces parallel to the weak vector $\vec{\nu}$,
they must appear in line in the direction of $\vec{\nu}$.

%%%%%%%%%%%%%%%%%%%%%%%%%%
\begin{figure}
\begin{center}
\includegraphics[width=8cm]{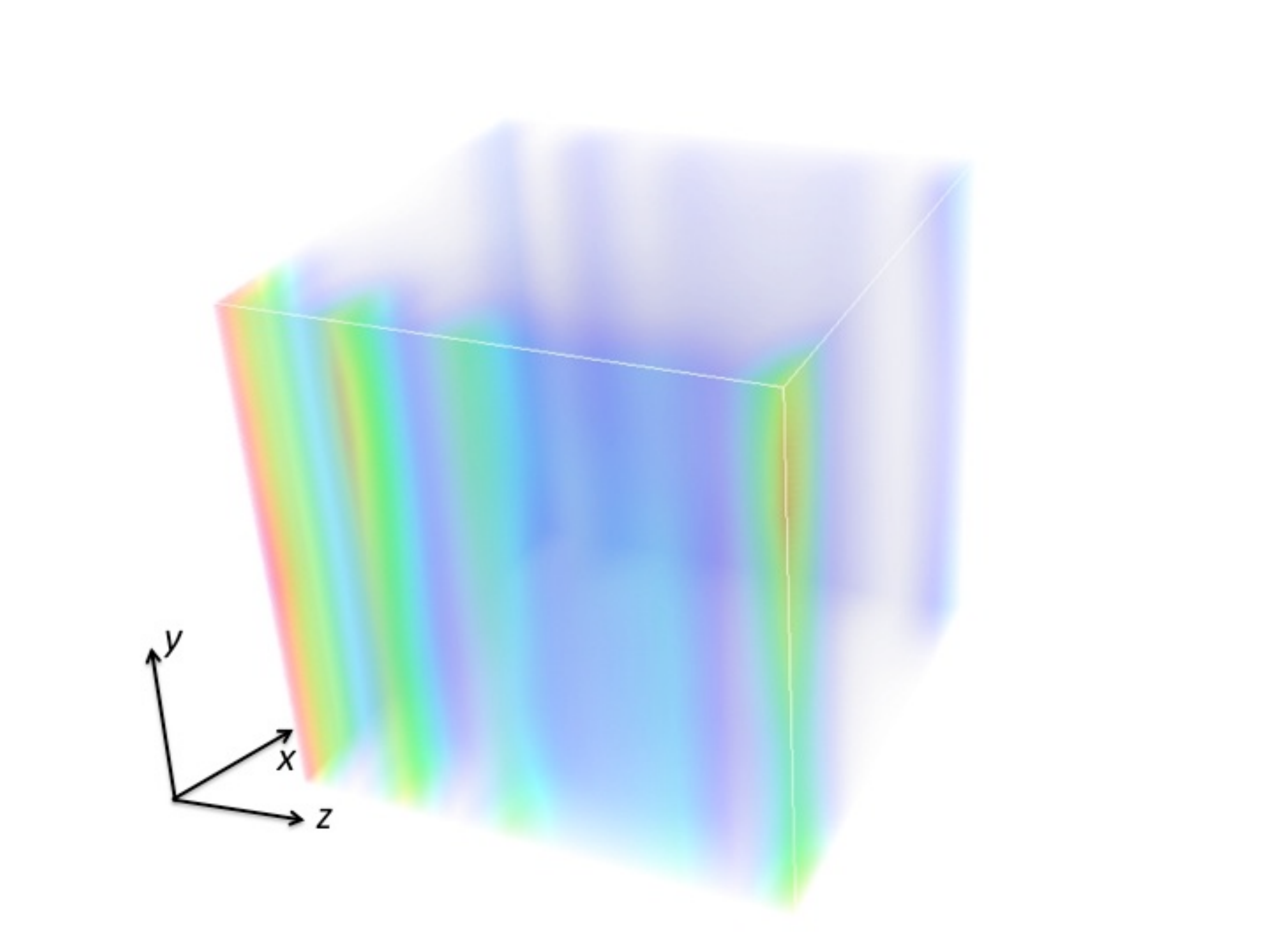}
\includegraphics[width=8cm]{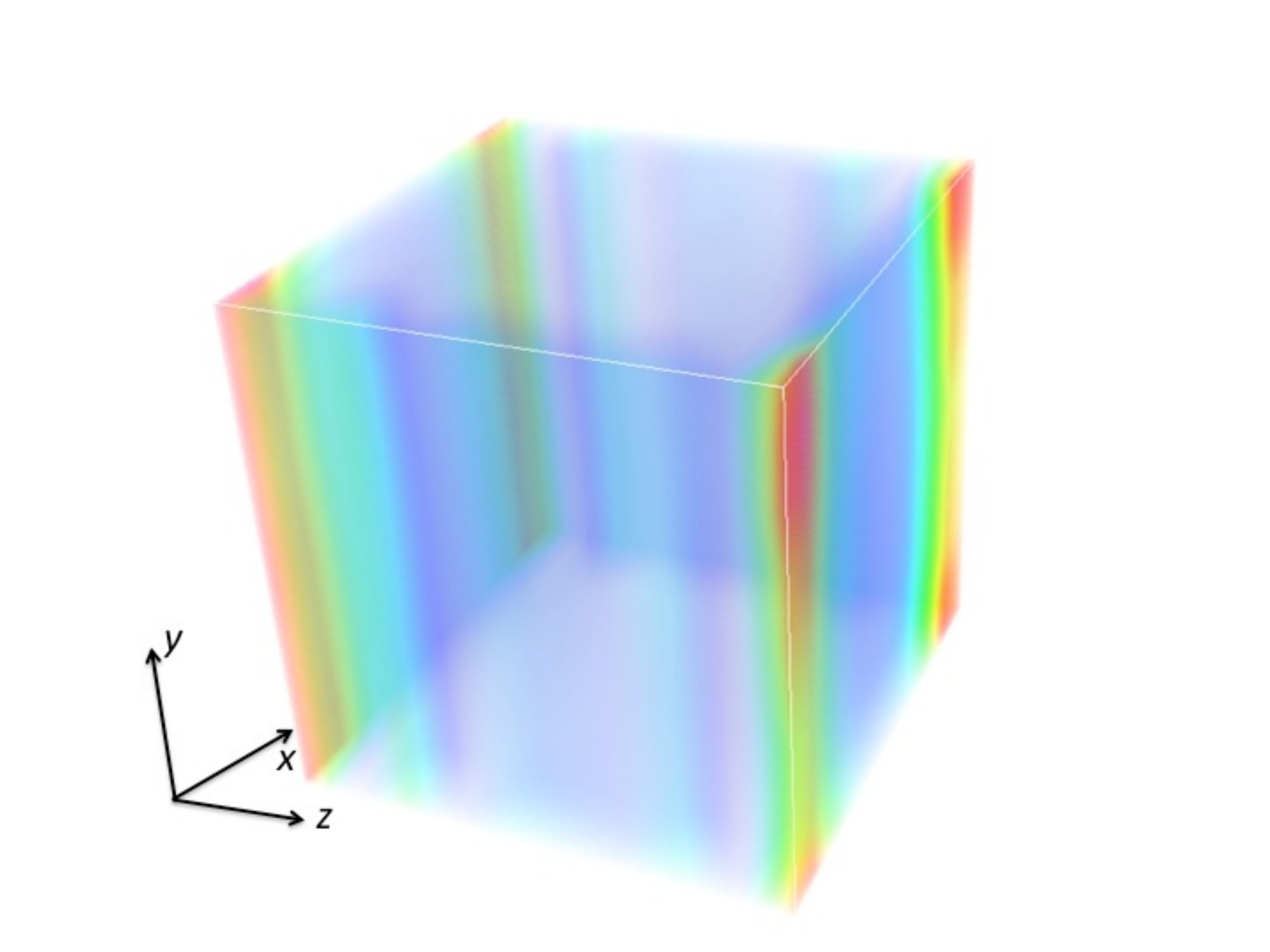}
\end{center}
\caption{Surface wave function in the presence of disorder.
Comparison of the WTI and STI cases:
$m_{2z}/m_{2\parallel} = 0.2$ (upper) vs. $m_{2z}/m_{2\parallel} = 0.3$ (lower).
Here, the simulation is done for a system of size,
$N_x \times N_y \times N_z = 10 \times 10 \times 10$;
{\it i.e.}, $N_z$ is even.}
\label{wf_disorder}
\end{figure}
%%%%%%%%%%%%%%%%%%%%%%%%%%

\subsection{Effects of disorder}

Let us comment here on the robustness of the surface states 
discussed in the previous subsections
against disorder.
A motivation for this is that
since disorder leads generally to repulsion of the energy levels,
one naturally questions whether
the finite-size effects discussed so far are still meaningful
when the size gap is perturbed by the effects of level repulsion by disorder.
The effects of disorder is taken into account by introducing
a random potential $V(\bm r)$, which obeys a uniform distribution
in the period $[-W/2, W/2]$ at each site $\bm r$ of the cubic lattice,
{\it i.e.}, a scalar random potential,
$\propto \bm 1$ in the real and orbital spin space,
which is also cite-diagonal:
\begin{equation}
V = \sum_{\bm r} V(\bm r) \bm 1 \otimes |\bm r\rangle \langle\bm r|,\ \
V(\bm r) \in [-W/2, W/2]
\label{random}
\end{equation}
is added to the tight-binding Hamiltonian (\ref{H_bulk})
represented in the real space. 
In Eq. (\ref{random}) the summation over $\bm r$
should be taken over all the lattice sites on the cubic lattice,
$\bm r = (x,y,z)$ with $x=1,2,\cdots, N_x$, $y=1,2,\cdots, N_y$
and $z=1,2,\cdots, N_z$.
In the actual computation 
%the strength $W$ of disorder is set to be unity.
we set $W=1$, $m_0 = -1$,
$A_\perp = A_\parallel =1$ in units of $m_{2\parallel}$
(which is set to be unity).

In FIG. \ref{wf_disorder}
plots similar to
FIG. \ref{wf0100_even}, FIG. \ref{wf0100_odd} and FIG. \ref{wf1000}
performed in the presence of disorder are shown.
%In the two upper panels (WTI case)
In the upper panel (WTI case)
the surface wave function is localized mainly %takane
on one facet of the prism.
This is contrasting to the clean cases (FIG. \ref{wf0100_even}, FIG. \ref{wf0100_odd})
and to the STI case (lower),
in which the surface state is extended over all the four facets of the prism.
The stripe-shaped structure is also still visible, indicating that
the surface wave functions of a specific shape discussed in the previous subsection
possess some robustness against disorder.

%%%%%%%%%%%%%%%%%%%%%%%%%%
\begin{figure}
\begin{center}
\includegraphics[width=7cm]{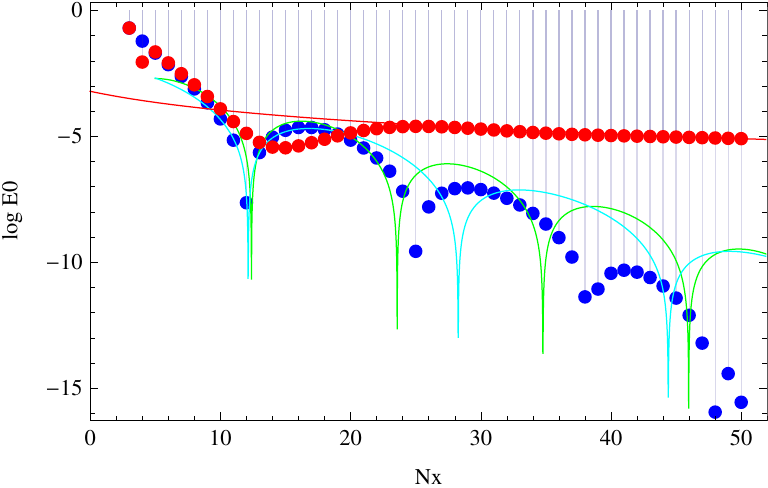}
\end{center}
\caption{Finite-size energy gap
in the rectangular-prism geometry
plotted as a function of the ``width'' $N_x$.
Comparison between the
WTI (blue points) and STI (red points) regimes in the case of prism thickness $N_z$ odd
($N_z = 9$).
The logarithm of the energy gap $E_0$ is plotted vs. $N_x$ 
for demonstrating that $E_0 = E_0 (N_x)$ decays exponentially,
showing actually  an exponentially damped oscillation
in the WTI phase.
The corresponding solutions of Eq. (\ref{rho12}) 
are a pair of complex numbers
(see main text for details).
The model parameters employed are also given there.}
\label{Nz_odd}
\end{figure}
%%%%%%%%%%%%%%%%%%%%%%%%%%

%%%%%%%%%%%%%%%%%%%%%%%%%%
\begin{figure}
\begin{center}
\includegraphics[width=7cm]{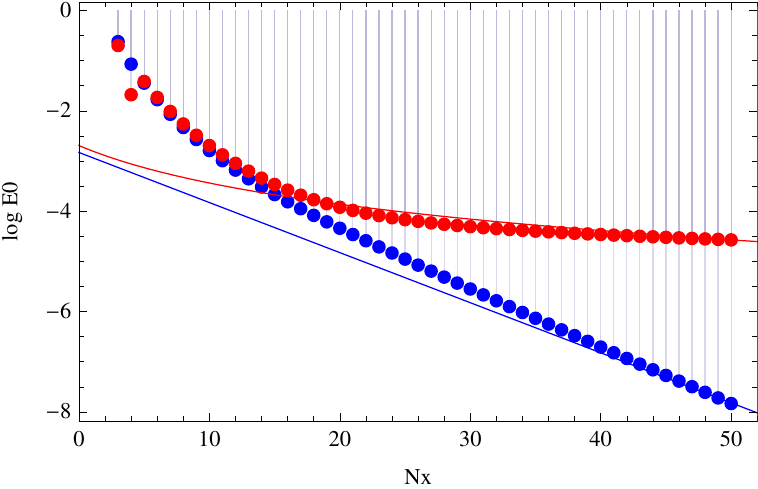}
\end{center}
\caption{Size dependence of $E_0 = E_0 (N_x)$  in the case of
$N_z$ odd ($N_z = 9$). 
A plot similar to FIG. \ref{Nz_odd}
but in the case of model parameters, yielding as solutions for
$\rho$ in Eq. (\ref{rho12}),
two real solutions
given in the main text.
The data points for the WTI and STI cases
are shown, respectively, in blue and in red.}
\label{Nz_odd2}
\end{figure}
%%%%%%%%%%%%%%%%%%%%%%%%%%

%%%%%%%%%%%%%%%%%%%%%%%%%%
\begin{figure}
\begin{center}
\includegraphics[width=7cm]{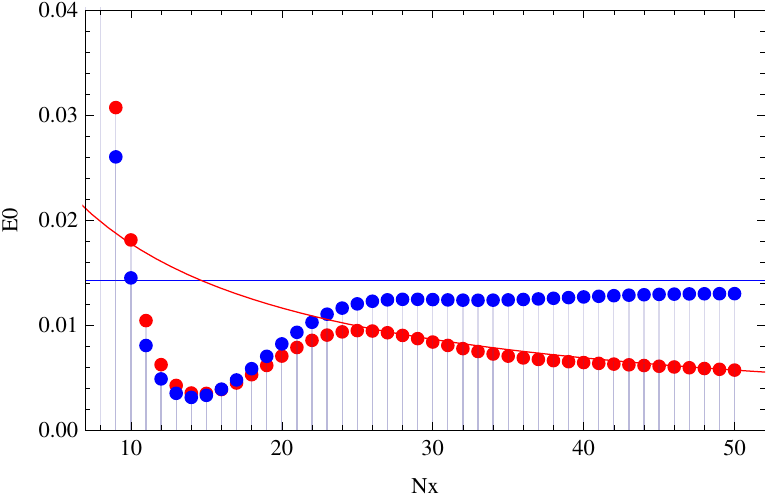}
\end{center}
\caption{Size dependence of $E_0 = E_0 (N_x)$  in the case of
$N_z$ even ($N_z = 10$).
The mass and velocity parameters are chosen to be
the same as in the case of FIG. \ref{Nz_odd}. 
Here, the vertical axis for $E_0$ is
in the linear scale. 
For $N_z$ even $E_0 (N_x)$ shows at most a power-law decay,
whether the system is
in the STI or WTI phase (see Table \ref{abc}).
The data points for the WTI and STI cases
are as before shown, respectively, as blue and red filled circles.}
\label{Nz_even}
\end{figure}
%%%%%%%%%%%%%%%%%%%%%%%%%%

\subsection{STI more gapped than WTI !?}

We finally discuss the $N_x$-dependence of the size gap.
As shown in Table \ref{abc},
there are three different types of behaviors
in the $N_x$-dependence of the size gap,
each corresponding to the three different gap-opening mechanisms
we have highlighted in this paper.
Here, let us focus again 
({\it cf.} FIG. \ref{gap_Nz})
on the (red) line $m_0/m_{2\parallel}=-1$ in the phase diagram (FIG. \ref{phase_diagram})
slightly above and below the phase boundary at $m_{2\perp}/m_0= -1/4$,
and compare the STI: (1,000) and WTI-A: (0,001) phases.
In the following demonstrations 
(FIG. \ref{Nz_odd}, FIG. \ref{Nz_odd2}, FIG. \ref{Nz_even}), 
however,
we use slightly different set of parameters
inspired by the corresponding material parameters of Bi$_2$Se$_3$,
\cite{Liu_PRB}
but focus on the same phase boundary between STI and WTI-A.
Here, the tight-binding parameters are specially adjusted
\cite{Ebi}
to reproduce the band structure in the vicinity of $Z$-point
obtained by the first-principle calculation.
The employed parameters are given explicitly as
\begin{eqnarray}
m_0 = -0.1,\ \
&& m_{2z} \equiv m_{2\perp} = 0.1,\ \ 
m_{2\parallel} = 1,
\nonumber \\
&& A_{z}\equiv A_{\perp} = 0.1,\ \ 
A_{\parallel} = 0.3.
\label{param_exp}
\end{eqnarray}
Here, the parameters are
normalized in units of $m_{2\parallel} \simeq$ 2.60 eV.
This set of parameters corresponds to the case of the STI phase.
To achieve a weak phase
we modify
the value of $m_{2\perp}$ in Eq. (\ref{param_exp})
as
$m_{2\perp} \rightarrow 0.01$.
This indeed falls on the WTI-A phase in FIG. 1.
The spectrum of the {\it strong} phase is ``gapped'', showing a finite-size energy gap
due to spin-to-surface locking, which decays only algebraically,
$E_0 \sim (N_z+N_x)^{-1} \neq 0$.
In the {\it weak} phase, 
and in the case of $N_z$ odd considered here,
the spectrum is ``gapless'', decaying exponentially
as a function of the distance $\sim N_x$ 
between the two {\it ideally} gapless patches
($\log E_0 \propto -N_x$, $E_0 \simeq 0$).
This is indeed a comparison of the cases (b) and (c) in Table \ref{abc}.
In FIG. \ref{Nz_odd},
the logarithm of the energy gap $E_0$ is plotted vs. $N_x$ 
taking into account such an expected exponential decay in the WTI-A phase.
But here, a systematic deviation from a simple exponential decay can be 
clearly seen, implying that this is rather a damped oscillation.

%\textcolor{blue}{
As mentioned in Appendix B,
the magnitude of the finite-size energy gap in the slab
is directly related to the
(complex) penetration depth of the surface wave function, 
or $\rho_{1,2}$ given in Eq. (\ref{rho12}).
One can indeed verify,
%\textcolor{blue}{
\begin{equation}
E_0 (N_x) \propto \left| \rho_1^{N_x +1} -\rho_2^{N_x +1} \right|.
\label{gap_rho}
\end{equation}
Recall that in the WTI-A phase considered here
two Dirac cones,
one at $\bm k_1 = (0,0)$ and 
the other at $\bm k_2 = (0,\pi)$,
are well grounded
on the $\hat{\bm x}$-surfaces.
The corresponding surface wave functions exhibit
different penetration depths at each Dirac point,
which are specified by Eq. (\ref{rho12}).
The solutions of Eq. (\ref{rho12}) at $\bm k_1 = (0,0)$ are
\begin{equation}
\rho = \rho_{1,2} (\bm k_1) \simeq 0.826 \pm 0.238\ i,
\label{rho12_0}
\end{equation}
while they are given by
\begin{equation}
\rho = \rho_{1,2} (\bm k_2) \simeq 0.843 \pm 0.166\ i,
\label{rho12_pi}
\end{equation}
at $\bm k_2 = (0,\pi)$,
{\it i.e.},
in the two cases,
they become a pair of complex numbers.
In the slab, 
the finite-size energy gap 
is $\bm k = (k_z, k_x)$-resolved;
$E_0 = E_0 (\bm k)$,
simply the minimal value of which determines the actual magnitude of
the finite-size energy gap.
In the case of rectangular prism,
contributions from $\bm k = \bm k_1$ and from $\bm k = \bm k_2$
are superposed to cope with the boundary condition.
Notice also that here
the surface wave function 
at $\bm k = \bm k_1$ and at $\bm k = \bm k_2$
are both oscillatory [Eqs. (\ref{rho12_0}) and (\ref{rho12_pi})].
These two features combine to give the oscillatory pattern
of $\log E_0$ in the WTI case in FIG. \ref{Nz_odd}.
In the figure,
two ``theoretical'' curves for $\log E_0$
are shown in solid curves for comparison,
not showing a quantitative agreement with the actual data.
The two curves correspond to the finite-size energy gap 
given as in Eq. (\ref{gap_rho})
at $\bm k = \bm k_1$ (green) and at $\bm k = \bm k_2$ (cyan)
estimated under the hypothesis that the system is slab-shaped.
The actual $N_x$-dependence of $\log E_0$ is somewhat in between.

FIG. \ref{Nz_odd2} is a plot similar to FIG. \ref{Nz_odd},
making the same comparison of the STI and WTI-A phases for the same $N_z$ odd case
except that
the model parameters are slightly modified from Eq. (\ref{param_exp}).
We replace one of the velocity parameters $A_{2\parallel}$
with $A_{2x} =0.7$, leaving $A_{2y}=0.3$ (the same value as before).
This replacement makes the corresponding solutions of Eq. (\ref{rho12})
two real solutions, indicating that the surface wave function 
exhibits a simple exponential decay.
In the WTI-A phase, we have chosen as before
$m_{2\perp} = 0.01$.
The behavior of $\log E_0$ in the WTI case is qualitatively different from the
previous case.
At the two Dirac points, $\bm k = \bm k_1$ and $\bm k = \bm k_2$,
in the WTI phase,
the solutions of Eq. (\ref{rho12}) are
\begin{equation}
\rho_1 (\bm k_1) \simeq 0.821,\ 
\rho_{2} (\bm k_1) \simeq 0.587
\end{equation}
at $\bm k_1 = (0,0)$, 
while they are given by
\begin{equation}
\rho_1 (\bm k_2) \simeq 0.905,\ 
\rho_{2} (\bm k_2) \simeq 0.532
\end{equation}
at $\bm k_2 = (0,\pi)$.
The actual magnitude of the size gap is determined by the largest value of
$\rho_{1,2}$, which is the value of $\rho_{1}$ at $\bm k = \bm k_2$.
Indeed,
the actual $N_x$-dependence of $\log E_0$ approaches to this scaling behavior
($E_0 \propto \rho_1 (\bm k_2)^{N_x}$, shown in a solid straight line in FIG. \ref{Nz_odd2})
for large enough $N_x$.

Through these two examples 
we can convince ourselves that 
in this configuration imposed by the combination of the prism geometry and a specific
choice of the weak vector,
which can be achieved by adjusting the direction of crystal growth direction with respect to the prism,
the strong topological insulator is qualitatively more gapped than 
a weak topological insulator.

In the last figure, FIG. \ref{Nz_even},
we make a comparison between the cases (a) and (c) in Table \ref{abc},
in contrast to the previous plots, the ones in
FIG. \ref{Nz_odd}, FIG. \ref{Nz_odd2}.
The model parameters 
are the same as in FIG. \ref{Nz_odd},
but here the number $N_z$ of stacking layers is even ($N_z =10$).
In the STI case the size gap shows a power law decay,
$E_0 \sim (N_z+N_x)^{-1} \neq 0$,
due to spin-to-surface locking.
In the WTI-A phase, the size gap implied by Eq. (\ref{gap_kp})
does not scale as a function of $N_x$, but
given simply by
\begin{equation}
E_0 = {A_z \over 2(N_z +1)}\pi = 0.1 \times {\pi \over 22}.
\label{E0_cst}
\end{equation}
In FIG. \ref{Nz_even}
this value is indicated as a horizontal grid line (in blue).
For sufficiently large value of $N_x$
the data looks almost constant
at a value not much far from the one of Eq. (\ref{E0_cst}).

We have seen so far that from the viewpoint of the scaling behavior of finite-size energy gap, 
{\it the statue of the strong and weak phases could be reversed}.
Here, to illustrate this feature
we have considered only a very representative range of parameters,
but the same feature is generic to the vicinity of transitions
between STI and WTI phases with a suitable choice of the surface directions 
and the number of quintuple layers.

\section{Conclusions}

We have studied the finite-size energy gap
in 3D weak and strong topological insulators.
Employing the standard Wilson-Dirac type effective model, we have developed
both numerical and analytical considerations.
It has been demonstrated that
anisotropy of the model and the geometry of the system
are among other model parameters crucial elements
for determining the qualitative nature of the finite-size energy gap.
The two elements manifest in a correlated manner.
The weak topological insulator (WTI) has a specific property of
(i) expelling the gapless surface state from surfaces normal to its
weak vector $\vec{\nu}$ ($\simeq$ weak indices), {\it i.e.}, no Dirac cone on the surface
normal to $\vec{\nu}$,
(ii) but on surfaces parallel to the weak vector, it bears two Dirac cones
[more Dirac cones than a strong topological insulator (STI)].
We have seen in this paper through the study of finite-size effects
that these two, seemingly competing characteristics of the WTI
operate, in fact, in a cooperative way
({\it c.f.}, $\bm k\cdot\bm p$-description of the surface state in the WTI phase;
Sec. IV-B).
The condition of no Dirac cone on the side normal to $\vec{\nu}$
imposes the relative orientation of the two Dirac cones 
on the side parallel to $\vec{\nu}$.
The weak indices are also much related to the anisotropy of the model
parameters.
To encompass different scaling behaviors of the
finite-size energy gap,
we have manipulated the weak indices
by varying the model parameters,
guided by the phase diagram shown in FIG. \ref{phase_diagram}.

Spin-to-surface locking is a characteristic feature of the 
topological insulator surface state,
operational both in the WTI and STI phases,
leading also to a finite-size energy gap that exhibits a specific power-law decay
as a function of the system's linear dimension. 
Clearly, this is more relevant than a usual exponential decay associated with 
the overlap of two surface wave functions, 
{\it e.g.}, sitting on the opposing sides of the slab geometry.
By its nature
the finite-size energy gap due to spin-to-surface locking is
not effective in the slab, but effective in the
prism-shaped geometry.
In the prism-shaped WTI samples,
the interplay of these three ingredients;
the weak vector, the spin-to-surface locking
and the rectangular-prism geometry
leads to intricate finite-size effects,
depending on the model parameters.
Three different gap opening mechanisms pointed out in this paper:
(i) mixing of the surface wave functions,
(ii) spin-to-surface locking, and
(iii) commensurability with the boundary condition,
are all effective in determining
the intricate size dependence of the energy gap 
in the rectangular-prism geometry.

\acknowledgments
The authors
acknowledge Keith Slevin, Koji Kobayashi,
Kazuto Ebihara, Keiji Yada and Ai Yamakage
for useful discussions. 
%on the roles of anisotropic mass parameters in the spectrum of TI thin films.
KI, YT and TO
are supported by KAKENHI; KI by the ``Topological Quantum Phenomena'' [No. 23103511], 
YT and TO by Grant-in-Aid for Scientific Research (C) [Nos. 24540375, 23540376].

\appendix

%\section{Determination of the phase diagram}

%%%%%%%%%%%%%%%%%%%%%%%%%%
\begin{figure}
\begin{center}
\includegraphics[width=7cm]{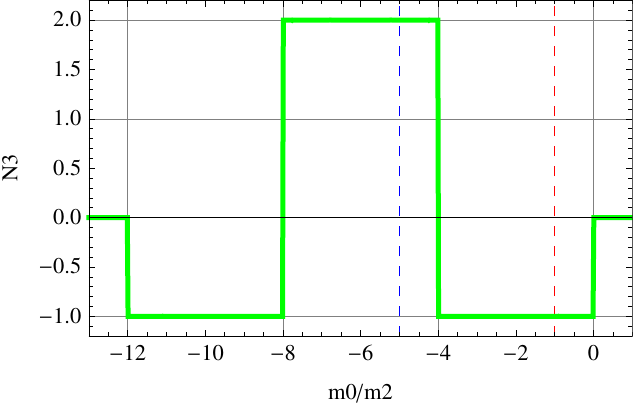}
\includegraphics[width=7cm]{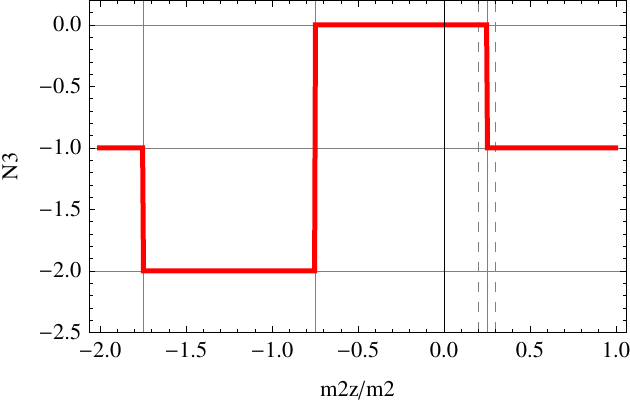}
\includegraphics[width=7cm]{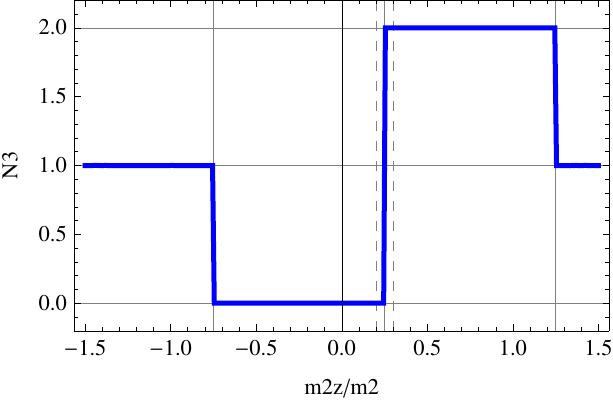}
\end{center}
\caption{The winding number $N_3$ [given in Eq. (\ref{N3})],
evaluated
on a horizontal or vertical line in FIG. 1.
In the first panel,
$m_{2\perp}/m_{2\parallel}$ is fixed at the isotropic point
($m_{2\perp}/m_{2\parallel}=1$), with $m_0/m_{2\parallel}$ being varied,
while in the remaining panels
$m_0/m_{2\parallel}$ is at fixed $m_0/m_{2\parallel} = -1$ (second panel)
and at $m_0/m_{2\parallel} = -5$ (third panel).
The lines are shown in the same color
in the phase diagram (see FIG. 1).
%Dashed colored lines indicate the mutual relation among the three panels.
%Dashed black lines correspond to $m_{2\perp}/m_{2\parallel} = 0.2$ and $m_{2\perp}/m_{2\parallel} = 0.3$.
}
\label{winding}
\end{figure}
%%%%%%%%%%%%%%%%%%%%%%%%%%

\section{Topological numbers}

Notice that our
model specified by Eqs. (\ref{H_bulk}), (\ref{mass_k})
has inversion symmetry.
This allows us 
to find the strong and weak $\mathbb Z_2$-indices
with the use of Fu-Kane's formula.
\cite{FuKane}
Here, we mention that in the specific case of
$\epsilon (\bm k) = 0$
(in most of the analyses in this paper we employ this condition for mathematical simplicity)
one can introduce a $\mathbb Z$-type winding number
$N_3$.
The strong index $\nu_0$ is related to $N_3$ as $\nu_0 = N_3 \mod 2$.

In terms of the periodic table
\cite{Zirnbauer, Schnyder_PRB, Kitaev_AIP, Schnyder_AIP, Ryu_NJP, TeoKane}
our starting bulk effective Hamiltonian (\ref{H_bulk})
falls on the class AII.
This class of models has the symmetry,
$\Theta^2=-1$, ${\cal C}^2=0$ and $\Gamma_5 =0$,
where
$\Theta$, $\cal C$ and $\Gamma_5$ represent, respectively,
the time-reversal, particle-hole and chiral symmetries,
and in this terminology
''$0$'' indicates that
the system does not possess that type of symmetry.
The periodic table says that 
class AII models are characterized by
$\mathbb{Z}_2$-type bulk topoloogical invariants
in 3D.
For the specific case of $\epsilon (\bm k) =0$
in our model,
the symmetry of the model is upgraded to the class DIII,
i.e.,
$\Theta^2=-1$, ${\cal C}^2=1$ and $\Gamma_5 =1$,
where for the specific Hamiltonian, Eq. (\ref{H_bulk}),
${\cal C}$ and $\Gamma_5$ are given by
${\cal C}=\sigma_y \tau_y K$ and
$\Gamma_5 =\tau_y$.
This symmetry class allows for $\mathbb{Z}$-type
bulk topological classification in 3D,
characterized by a $\mathbb{Z}$-type winding number $N_3$
to be defined below.

To construct the winding number $N_3$ explicitly,
let us first represent the bulk Hamiltonian (\ref{H_bulk}),
using an explicit matrix representations for the orbital Pauli matrices
$\tau_x$ and $\tau_y$
as
\begin{equation}
H_{\rm bulk} =
\left[
\begin{array}{cc}
0 & m(\bm k) - i P_\mu(\bm k)\sigma_\mu \\ 
m(\bm k) + i P_\mu(\bm k)\sigma_\mu & 0 
\end{array}
\right],
\end{equation}
where
we have introduced
$P_\mu (\bm k) = A_\mu \sin k_\mu$.
Dividing the Hamiltonian by (the magnitude of) its own eigenvalue $E(\bm k)$, 
one can also flatten the spectrum of the Hamiltonian as
\begin{equation}
\widetilde{H} (\bm k) ={H_{\rm bulk} (\bm k) \over |E(\bm k)|}=
\left[
\begin{array}{cc}
0 & Q (\bm k)\\ 
Q^\dagger (\bm k) & 0
\end{array}
\right],
\end{equation}
where
$E(\bm k) = \pm \sqrt{m(\bm k)^2 + P_\mu (\bm k)^2}$,
and
\begin{equation}
Q(\bm k) = {m(\bm k) - i P_\mu(\bm k)\sigma_\mu \over |E(\bm k)|}.
\end{equation}
Note that the matrix $Q$ defined above is
a $2\times2$ $\mathbb{SU}(2)$ matrix,
satisfying
$Q^\dagger Q=\bm 1$ and $\det Q=1$. 
Then, one can introduce 
an integral winding number $N_3$,\cite{spherical, Volovik1, Volovik2}
characterizing the mapping of the 3D Brillouin zone 
onto this $\mathbb{SU}(2)$ matrix as
\begin{equation}
N_3 = \frac{1}{24\pi^2}\int_{\rm BZ} d^3 k\ 
\epsilon_{\mu\nu\lambda}{\rm Tr}
\left[\Gamma_\mu \Gamma_\nu \Gamma_\lambda\right],
\label{N3}
\end{equation}
where
$\Gamma_\mu = Q^\dagger \partial_{k_\mu} Q$.
The integration should be done over the entire 3D Brillouin zone.
We have evaluated this winding number numerically
over the entire range of parameters shown in FIG. \ref{phase_diagram}
to verify that
\begin{equation}
\nu_0 = N_3 \mod 2
\end{equation}
indeed holds.
The explicit values of $N_3$ in the different STI and WTI phases
are also shown in FIG. \ref{phase_diagram}.
The same calculated value is also shown continuously
in FIG. \ref{winding}
as a function of a control parameter,
either $m_0/m_{2\parallel}$ or $m_{2\perp}/m_{2\parallel}$
on a few specific lines in FIG. \ref{phase_diagram}.

\section{Penetration of the surface wave function in the slab geometry}

To quantify the surface electronic state in the slab geometry,
let us concentrate on one surface of the slab.
Also, %for a later convenience, 
we choose this flat surface
normal to the $\hat{\bm x}$-direction.
To find the wave function
which is localized in the vicinity of the surface
we divide the bulk Hamiltonian (\ref{H_bulk})
into two parts:
\begin{equation}
H_{\rm bulk} (\bm k) = H_\parallel (\bm k_\parallel) + H_\perp (k_x),
\label{decomp_slab}
\end{equation}
where $\bm k_\parallel = (k_y, k_z)$ and
\begin{equation}
H_\parallel  (\bm k_\parallel) = \tau_x m_\parallel (\bm k_\parallel) + \tau_y (\sigma_y A_y \sin k_y + \sigma_z A_z \sin k_z),
\label{H_2D}
\end{equation}
with $m_\parallel (\bm k_\parallel)$ defined as
\begin{equation}
m_\parallel (\bm k) = m_0 + 2 m_{2x} + 2 m_{2y} (1-\cos k_y)+2 m_{2z} (1-\cos k_z),
\label{m_2D}
\end{equation}
and 
\begin{equation}
H_\perp (k_x) = -2\tau_x m_{2x} \cos k_x + \tau_y \sigma_x A_x \sin k_x.
\label{H_perp_A1}
\end{equation}
This and the following procedure is in parallel with the case in which
we deal with the continuum model,
a more standard situation in the context of $\bm k\cdot\bm p$ approximation,
%\textcolor{blue}{
discussed in Appendix C,
but here we solve the lattice model directly without taking the continuum limit.
\cite{Mao, HgTe_JPSJ}
Physically the decomposition (\ref{decomp_slab}) 
is based on the picture that
each ($y,z$)-plane described by $H_\parallel (\bm k_\parallel)$ is coupled by $H_\perp (k_x)$
to the neighboring layers.
In the present geometry,
$\bm k_\parallel = (k_y, k_z)$ is a good quantum number.
%Note that this ``stacking'' in the direction normal to the surface is simply geometrical, and has nothing to do with the physical stacking in the direction of $\vec{\nu}$.
Here, we assume that the system is extended in the half space: $x\ge 1$,
and impose a boundary condition: $\psi (x=0)=0$.
A surface solution in such a geometry can be constructed 
by composing a linear combination of base solutions of the form,
$\psi (x) = \rho^x \psi_0$ ($|\rho|<1$).
For such damped (instead of plane-wave) solutions, 
Eq. (\ref{H_perp_A1}) modifies to
\begin{equation}
H_\perp (\rho) = -2\tau_x m_{2x} {\rho + \rho^{-1} \over 2} + \tau_y \sigma_x A_x {\rho - \rho^{-1} \over 2i}.
\label{H_perp_A2}
\end{equation}
In the surface energy spectrum $E=E (\bm k_\parallel)$,
protected gapless Dirac points can appear at either of the four TRIM:
$\bm k_{\rm TRIM} = (0,0), (\pi,0), (0,\pi), (\pi,\pi)$.
%Later, when the system is put into the nanowire geometry, the wave function of the surface state is faced with an additional boundary condition at each edge of the surface.
%The wave functions $\psi (\bm k_\parallel)$ at each valley of the WTI phase combine to cope with such boundary conditions.
%Thus, the structure of the wave functions at the surface TRIM (the Dirac points) is most relevant to determining the behavior of finite-size energy gap.
At such TRIM of the surface BZ, the hopping terms in $H_\parallel (\bm k_\parallel)$
become inert;
\begin{equation}
H_\parallel (\bm k_\parallel = \bm k_{\rm TRIM})
= \tau_x m_\parallel (\bm k_{\rm TRIM}).
\end{equation}
This significantly simplifies the derivation of $\psi (\bm k_\parallel)$ 
at $\bm k_\parallel = \bm k_{\rm TRIM}$.
%%%%%%%%%%%%%%
Notice also that
Eq. (\ref{H_2D}) with (\ref{m_2D}) can be regarded as a lattice Hamiltonian for
a 2D $\mathbb Z_2$ TI with an effective mass parameter $m_{\rm 2D} = m_0 + 2 m_{2x}$.
$m_{\rm 2D} <0$ ($m_{\rm 2D} >0$)
corresponds, respectively, to the
non-trivial ($\nu=1$) vs. trivial ($\nu=0$) phases, where $\nu$ is the 2D $\mathbb Z_2$ index.
A situation described by this couple of equations
realizes in the limit $N_x \rightarrow 1$.

Let us construct the surface wave function,
\begin{equation}
\psi (\bm k_\parallel, x) = \rho^x \psi_0 (\bm k_\parallel),
\end{equation} 
explicitly at $\bm k_\parallel = \bm k_{\rm TRIM}$.
At TRIM, $\psi_0 (\bm k_\parallel)$ satisfies,
\begin{equation}
H_{\rm bulk} \psi_0 =
\left[ \tau_x m_\parallel (\bm k_{\rm TRIM}) + H_\perp (\rho) \right] \psi_0 = \bm 0,
\end{equation} 
i.e., $\psi_0$ is a zero-energy eigenstate of
\begin{equation}
\tau_x  H_{\rm bulk} = m_\parallel (\bm k_{\rm TRIM}) 
- m_{2x} (\rho + \rho^{-1}) + \tau_z \sigma_x {A_x \over 2}(\rho -\rho^{-1}).
\label{reduced}
\end{equation}
Similarly to the case of the continuum model 
%\textcolor{blue}{
(see Appendix C), 
this zero-energy condition is proven to be necessary \cite{spherical} 
for constructing a surface solution compatible with the boundary condition 
at $x=0$ in the form of Eq. (\ref{psi_rho}).
Clearly, any of the four simultaneous eigenstates of $\tau_z$ and $\sigma_x$,
$\psi_{\pm\pm}= |\tau_z \pm\rangle|\sigma_z \pm\rangle$,
is an eigenstate of the reduced operator (\ref{reduced}).
Then the zero-energy condition can be used, in turn,
to determine $\rho$ as
\begin{equation}
\rho = {m_\parallel \pm \sqrt{m_\parallel^2 -4(m_{2x}^2 -A_x^2/4)}\over 2(m_{2x}\pm A_x/2)}
\equiv \rho_{1,2}
\label{rho12}
\end{equation}
where
\begin{equation}
m_\parallel = m_\parallel (\bm k_{\rm TRIM}) \equiv m_0 (\bm k_{\rm TRIM}) + 2 m_{2x}.
\end{equation}
Here,
$m_0 (\bm k_{\rm TRIM})$ represents
the magnitude of bulk energy gap at $\bm k = \bm k_{\rm TRIM}$.
In Eq. (\ref{rho12}) 
the meaning of two double signs 
may need some explanation;
the one in the numerator is arbitrary, each choice
corresponding to $\rho_{1,2}$.
The one in the denominator represents
$+$ for $\psi_0 =\psi_{++}$ and $\psi_{--}$,
whereas,
the same sign represents $-$ for $\psi_0 =\psi_{+-}$ and $\psi_{-+}$.
The structure of Eq. (\ref{reduced}) 
with the understanding that $\tau_z \sigma_x = \pm 1$
indicates that
if $\rho$ satisfies the zero-energy condition, so does $\rho^{-1}$.
With a suitable choice of $\psi_0$, 
satisfying both $|\rho_1|<1$ and $|\rho_2|<1$,
the surface solution can be constructed as
%\textcolor{blue}{
\begin{equation}
\psi (x) = (\rho_1^x - \rho_2^x) \psi_0.
\label{psi_rho}
\end{equation}

In a separate paper
\footnote{K.-I. Imura and Y. Takane, to appear.}
we study in detail 
various aspects of the finite-size effects in a slab-shaped sample.
The magnitude of the finite-size energy gap in the slab
is determined by the
overlap of the two surface wave functions sitting on the opposing sides of the
slab.
It is, therefore, naturally expected that
the magnitude of the gap (in a slab of width $N_x$)
is essentially determined by
the penetration depth, or
the amplitude of the wave function (\ref{psi_rho})
at the depth of $x=N_x$.
Here, in this model one can verify that
the correlation of theses two quantities is a bit stronger than this.
The magnitude of the size energy gap $E_0 (N_x)$
is indeed directly proportional to $|\psi (N_x)|$
as given in Eq. (\ref{gap_rho}).

\section{Derivation of the effective surface Hamiltonian in the cylinder geometry}

To find the surface effective Hamiltonian on the cylinder
in the spirit of $\bm k \cdot \bm p$ approximation,
\cite{Liu_PRB, Shen_NJP, k2}
one first divides the bulk 3D effective Hamiltonian (\ref{H_Gamma})
into two parts;
one perpendicular, the other parallel to the cylindrical surface:
\begin{equation}
\label{decomp}
H = H_\perp (p_r) + H_\parallel (p_\phi, p_z),
\end{equation}
where $H_\perp = H|_{p_\phi=p_z=0}$, and
$p_r = -i \partial /\partial r$.
$H_\perp$ and $H_\parallel$ read explicitly,
\begin{eqnarray}
H_\perp &=& m_\perp \tau_x + A p_r \tau_y (\sigma_x \cos \phi + \sigma_y \sin \phi)
\nonumber \\
&=& \tau_x \left[ m_\perp + i A p_r \tau_z  \hat{\bm r} \cdot \bm \sigma \right],
\label{H_perp1} \\
H_\parallel &=& m_\parallel \tau_x + A \tau_y \left[ p_\phi (-\sin \phi \sigma_x + \cos \phi \sigma_y) + p_z \sigma_z \right]
\nonumber \\
&=& m_ \parallel \tau_x + A \tau_y \left[ p_\phi \hat{\bm \phi} \cdot \bm \sigma + p_z \sigma_z \right],
\end{eqnarray}
where
\begin{equation}
m_\perp = m_0 - m_2
\left[ {\partial^2 \over \partial r^2} + {1 \over r}{\partial \over \partial r} \right],
\end{equation}
and
$m_\parallel = m_2 (p_\phi^2 + p_z^2)$,
with
\begin{equation}
p_\phi = -i {1 \over r}{\partial \over \partial \phi},\ \
p_z = -i {\partial \over \partial z}.
\end{equation}
We have also introduced
$\hat{\bm r}=(\cos \phi, \sin \phi)$, and
$\hat{\bm \phi}=(-\sin \phi, \cos \phi)$.

We then consider a solution of the eigenvalue equation,
\begin{equation}
H_\perp |\psi_\perp \rangle\rangle = E_\perp |\psi_\perp \rangle\rangle
\end{equation}
of the form,
$\psi_\perp \sim e^{\kappa (r-R)}$,
i.e., we set $p_r = -i\kappa$ ($\kappa >0$) in Eq. (\ref{H_perp1}).
$E_\perp$ is the value of energy eigenvalue at the Dirac point.
In order to cope with the boundary condition
$| \psi_\perp \rangle_{r=R} = \bm 0$
on the surface of the cylinder,
one can verify that this must be zero ($E_\perp =0$).
\cite{spherical, k2}
This implies,
\begin{equation}
\tau_x H_\perp |\psi_\perp \rangle\rangle = \bm 0.
\label{eigen_0}
\end{equation}
Notice that in the second line of Eq. (\ref{H_perp1})
$\hat{\bm r} \cdot \bm \sigma$ can be diagonalized by pointing the real-spin spinor
in the direction of $\hat{\bm r}$ as Eqs. (\ref{r_dv}).
Then, one can satisfy Eq. (\ref{eigen_0})
by four simultaneous eigenstates of $\tau_y$ and $\hat{\bm r} \cdot \bm \sigma$,
i.e.,
\begin{equation}
|\psi_\perp \rangle\rangle = \rho (r)
|\tau_z \pm\rangle |\hat{\bm r} \pm\rangle_{\rm dv},
\label{spinor}
\end{equation}
if $\kappa$ is a solution of 
\begin{eqnarray}
E_\perp = m_\perp \pm A\kappa
= m_0 - m_2 \kappa^2 \pm A\kappa = 0.
\label{E_perp}
\end{eqnarray}
$|\hat{\bm r} \pm\rangle_{\rm dv}$ has been given in Eqs. (\ref{r_dv}).
The double sign in Eq. (\ref{E_perp})
signifies $+$ ($-$)
when the combination of two signs in
$|\tau_y \pm\rangle |\hat{\bm r} \pm\rangle$ in Eq. (\ref{spinor})
are the same (opposite).
One has to consider a linear combination of the eigenstates
of the form,
\begin{eqnarray}
\rho (r) \sim e^{\kappa_1 (r-R)} - e^{\kappa_2 (r-R)}
\label{rho}
\end{eqnarray}
where $\kappa_1$ and $\kappa_2$ are solutions of Eq. (\ref{E_perp})
with $E_\perp =0$, i.e.,
\begin{equation}
\kappa = {\pm A \pm \sqrt{A^2 + 4 m_0 m_2} \over 4 m_2} \equiv \kappa_{1,2},
\label{kappa12}
\end{equation}
where the double sign in front of $A$ corresponds to the one in Eq. (\ref{E_perp}).
The second one is arbitrary, each choice determining the subscript of $\kappa_{1,2}$.
Here, the surface state should be localized in the inner vicinity of the surface of the cylinder.
For that one needs a solution of the form of Eq. (\ref{rho}) with $\kappa_{1,2}$
whose real part {\it both} being positive.
This is in one-to-one correspondence with
\begin{itemize}
\item 
 the choice of $+$ sign in front of $A$ in Eq. (\ref{kappa12}),
{\it assuming that $A / m_2$ is positive},  \underline{and}
\item
the condition $m_0 m_2 <0$.
\end{itemize}
Thus,
the two basis solutions 
that span the subspace of the surface solutions of Eq. (\ref{bulk_eigen})
that are also compatible with the boundary condition
are identified as $|\hat{\bm r} \pm\rangle\rangle_{\rm dv}$,
introduced in Eqs. (\ref{r_dv2}).
For preciseness, we normalize
Eq. (\ref{rho}) as
\begin{equation}
\rho (r) = \sqrt{\kappa_1 \kappa_2 (\kappa_1 +\kappa_2) \over \pi R}
{e^{\kappa_1 (r-R)} - e^{\kappa_2 (r-R)} \over |\kappa_1 - \kappa_2|}.
\label{rho_n}
\end{equation}
Any surface solution $|\bm \alpha \rangle\rangle$ of Eq. (\ref{bulk_eigen}),
satisfying
\begin{equation}
H_\parallel |\bm \alpha \rangle\rangle = E |\bm \alpha \rangle\rangle,
\label{eigen_1}
\end{equation}
can be expressed as a linear combination of these two basis solutions as
\begin{equation}
|\bm \alpha \rangle\rangle =
\alpha_+
|\bm r + \rangle\rangle_{\rm dv}
+
\alpha_-
|\bm r - \rangle\rangle_{\rm dv},
\label{alpha_bulk2}
\end{equation}
or as in Eq. (\ref{alpha_bulk}).

Finally, following the prescription of the standard degenerate perturbation theory,
we consider the secular equation for Eq. (\ref{eigen_1}),
i.e.,
\begin{equation}
\left[
\begin{array}{cc}
\langle\langle \bm r +| H_\parallel |\bm r +\rangle\rangle
& \langle\langle \bm r +| H_\parallel |\bm r -\rangle\rangle
\\ 
\langle\langle \bm r -| H_\parallel |\bm r +\rangle\rangle & 
\langle\langle \bm r -| H_\parallel |\bm r -\rangle\rangle
\end{array}
\right]
\left[
\begin{array}{c}
\alpha_+ \\ \alpha_-
\end{array}
\right] 
= E
\left[
\begin{array}{c}
\alpha_+ \\ \alpha_-
\end{array}
\right],
\label{secular}
\end{equation}
where we have omitted the subscript ``dv'', for simplicity.
We define the coefficient matrix 
$\langle\langle \bm r\pm| H_\parallel |\bm r\pm\rangle\rangle$
in the secular equation Eq. (\ref{secular})
as the surface effective Hamiltonian $H_{\rm surf}$.
Noticing the relations such as
\begin{eqnarray}
\langle \hat{\bm r} \pm| \hat{\bm \phi} \cdot \bm \sigma |\hat{\bm r} \pm \rangle = \sigma_y,
\\
\langle \hat{\bm r} \pm| \sigma_z |\hat{\bm r} \pm \rangle = \sigma_x,
\end{eqnarray}
the explicit form of $H_{\rm surf}$ is found 
as given in Eq. (\ref{H_surf}).

%\section{Orbital-to-surface locking}

\bibliography{W1_r2_v4bis}

\end{document}